\documentstyle[12pt,graphicx]{article}

\newcommand{\la}{\langle}
\newcommand{\ra}{\rangle}

\begin{document}

\title{\bf The rational parts of one-loop QCD amplitudes III: The six-gluon case}

\author{Zhi-Guang Xiao\thanks{E-mail: zhgxiao@itp.ac.cn} $^{1,2}$,
Gang Yang\thanks{E-mail: yangg@itp.ac.cn} $^{2}$and Chuan-Jie
Zhu\thanks{E-mail: zhucj@itp.ac.cn} $^{2,3}$ }

\maketitle

\medskip\centerline{$^1$The Interdisciplinary Center of Theoretical
Studies, Chinese} \centerline{Academy of Sciences, P. O. Box 2735,
Beijing 100080, P.~R.~China}
\medskip
\centerline{$^2$Institute of Theoretical Physics, Chinese Academy
of Sciences} \centerline{P. O. Box 2735, Beijing 100080, P. R.
China}
\medskip
\centerline{$^3$Center of Mathematical Science, Zhejiang
University} \centerline{Hangzhou 310027, P. R. China}

\begin{abstract}

The rational parts of 6-gluon one-loop amplitudes with scalars
circulating in the loop are computed by using the newly developed
method for computing the rational parts directly from Feynman
integrals. We present the analytic results for the two MHV
helicity configurations: $(1^-2^+3^+4^-5^+6^+)$ and $
(1^-2^+3^-4^+5^+6^+)$, and the two NMHV  helicity configurations:
$ (1^-2^-3^+4^-5^+6^+)$ and $ (1^-2^+3^-4^+5^-6^+)$. Combined with
the previously computed results for the cut-constructible part,
our results are the last missing pieces for the complete partial
helicity amplitudes of the 6-gluon one-loop QCD amplitude.

\end{abstract}
\newpage

\section{Introduction}
In two previous papers \cite{xyzi,xyzii}, we developed a method of
computing the rational parts of one-loop amplitudes directly
from Feynman integrals and applied it to compute the rational
parts of the 5-gluon amplitudes. The result agrees with the
well-known result of Bern, Dixon and Kosower \cite{BDK} obtained
firstly by using string-inspired methods. In this paper we use our
method to compute explicitly the rational parts of the 6-gluon
amplitudes in QCD. For all the 8 independent helicity
configurations, we computed explicitly all cases except the all
plus helicity case, as this is a well-known finite amplitude
\cite{BCDK}. For the three helicity configurations which have
explicit results published \cite{Mahlon,BDKA,BDKB,BDKC}, we found
complete analytic agreement. For the remaining two MHV cases we
compared with the results of Berger, Bern, Dixon, Forde and Kosower
\cite{BDKE} obtained by using the bootstrap recursive relations.
We find complete agreement. The results obtained by Berger, Bern,
Dixon, Forde and Kosower in \cite{BDKE} are more general
all-multiplicity one-loop MHV amplitudes. Our method shows its
power by firstly obtaining also the analytic results for the
rational parts of the remaining 2 non-MHV helicity configurations of the
one-loop 6-gluon amplitude. Our method and the bootstrap recursive
approach complement each other quite nicely.

The constant effort to calculate higher-point one-loop amplitudes
in general and 6-point amplitude in particular lies in the
application to the forthcoming experimental program at CERN's
Large Hadron Collider (LHC), as there are lots of processes with
many particles as final states \cite{Salam}. We refer the reader
to \cite{xyzi} for a discussion and extensive references for the
recent efforts in computing the multi-particle one-loop amplitudes
and the recent developments inspired by twistor string theory
\cite{Witten,CSW}. Here we will concentrate on the 6-gluon
one-loop amplitude in QCD.

In principle any one-loop amplitude in QCD can be computed by
using Feynman diagrams and Feynman rules. The 4-parton amplitude
was computed in this way by computing the numerous Feynman diagrams
\cite{EllisSexton}. However using this brute-force method to
compute even the 5-gluon one-loop amplitude was quite a challenge.
Starting from 1987,  Bern and Kosower developed the method to
compute one-loop amplitudes by string theory \cite{BernKosower87}.
The 5-gluon one-loop amplitude was computed firstly in this way by
Bern, Dixon and Kosower \cite{BDK}. It turns out that the lessons
learnt from string theory are quite useful. In fact one can forget
about string theory and only keeps the tricks learnt in these
lessons \cite{BernReview,ChineseMagic,DixonReview}. Subsequently
other 5-parton amplitudes were later computed by using either
standard Feynman diagrammatic technique \cite{KST} or
supersymmetric decomposition and perturbative unitarity
\cite{BDKM}.

However the string theory (inspired) method is still not powerful
enough and the analytic computation of the 6-gluon one-loop
amplitude is quite a challenge. Only for special helicity configurations and
special models, some analytic results are known
\cite{BCDK,Mahlon,BinothYukawa}.

To make progress, it is a good strategy to decompose the QCD
amplitude into simpler ones by using the supersymmetric decomposition:
\begin{equation}
A^{QCD} = A^{N=4} - 4 A^{N=1~{\rm chiral}} + A^{N=0~{\rm or~
scalar}} ,
\end{equation}
where $A^{QCD}$ denotes an amplitude with only a gluon circulating
in the loop, $A^{N=4,1}$ have the full $N=4,1$ multiplets
circulating in the loop, and $A^{N=0}$ has only a complex scalar
in the loop.

By using the general properties of the one-loop amplitude, Bern,
Dunbar, Dixon and Kosower  proved that the supersymmetric
amplitude  $A^{N=4,1}$ are completely determined by 4-dimensional
unitarity \cite{BDDK}, i.e. the amplitude is completely
cut-constructible and the rational part is vanishing (see
\cite{xyzi} for more detail explanation). For MHV helicity
configurations, explicit results were obtained for $A^{N=4}$ in
\cite{BDDK}. The recent development of using MHV vertices to
compute one-loop amplitudes leads to many new results for the
cut-constructible part
\cite{BST,Cachazo,RecentTwistorBern,Rozali,BBST,Dunbar,BernA,Luo,
BCFW,BoFengA,BoFengC}. In particular, Bedford, Brandhuber,  Spence
and  Travaglini \cite{BST,BBST} applied the MHV vertices to
one-loop calculations. Britto, Buchbinder, Cachazo, Feng and
Mastrolia \cite{BoFengA, BoFengC, BoFengSix} developed an efficient
technique for evaluating the rational coefficients in an expansion
of the one-loop amplitude in terms of scalar box, triangle and
bubble integrals (the cut-constructible part, see \cite{xyzi} for
details). By using their technique, it is much easier to calculate
the coefficients of box integrals without doing  any integration.
Recently, Britto, Feng and Mastrolia completed the computation of
the cut-constructible terms for all the 6-gluon helicity
amplitudes \cite{BoFengSix}.

In order to complete the QCD calculation for the 6-gluon helicity
amplitudes,  the remaining challenge is to compute the rational
parts of the helicity amplitudes with scalars circulating in the
loop.

As we reviewed in \cite{xyzi}, there are various approaches
\cite{BDKX,BernMorgan,BDKY,AMST,BDKA, BDKB} to compute the
rational parts. In particular, Bern, Dixon and Kosower \cite{BDKA,
BDKB} developed the bootstrap recursive approach which has lead
to quite general results \cite{FordeKosower,BDKC,BDKE}. In this
paper we will use the approach as developed in \cite{xyzi} and
apply it to compute the rational parts of the 6-gluon one-loop
amplitudes in QCD (see also \cite{BDKE}) which are the last missing
pieces for the complete partial helicity amplitudes of the 6-gluon
one-loop QCD amplitude. The usefulness of this approach was
checked in \cite{xyzii} by reproducing (the rational parts of) the
well-known 5-gluon one-loop amplitudes of Bern, Dixon and Kosower
\cite{BDK}.  See \cite{BernReview} for a general review.

For the helicity configurations $(1^+2^+3^+4^+5^+6^+)$ and
$(1^-2^+3^+4^+5^+6^+)$, the cut-constructible parts are zero and
the rational parts were known already \cite{BCDK,Mahlon}. These
rational parts have also been derived by using the bootstrap
recursive method \cite{BDKC}. We also computed the rational part
of $(1^-2^+3^+4^+5^+6^+)$ by using our new method and found the
same analytic result.

For the so call ``split-helicity" configurations
$(1^-2^-3^+4^+5^+6^+)$ and $(1^-2^-3^-$ $4^+5^+6^+)$, the
computation of the rational parts is a recent achievement
\cite{BDKB,BDKC}. By using our method, we also computed the
rational parts for these two helicity configurations. The results
obtained agree with their results.
The proof is done by using
Mathematica by expressing all spinor products in terms of 12
independent spinor products  after explicitly solving the
equations from momenta conservation.  We will not give any details
in this paper.

A general strategy for computing the remaining 4 helicity
configurations was outlined in \cite{BDKA}. Explicit results for
the all-multiplicity MHV one-loop amplitudes, including the
remaining two MHV helicity configurations of the 6 gluons
$(1^-2^+3^-4^+5^+6^+)$ and $(1^-2^+3^+$ $4^-5^+6^+)$, were given
in \cite{BDKE}.  Independently we also obtained the explicit
results of the rational parts for these two MHV helicity
configurations of the 6 gluons, see Sects.~4 and 5.

Due to the appearance of the 3-mass triangle integrals in the
remaining 2 non-MHV helicity configurations $(1^-2^-3^+4^-5^+6^+)$
and $(1^-2^+3^-$ $4^+5^-6^+)$, the analytic results are much more
complicated. The results can be actually presented in a compact
form by exploiting the symmetry of the amplitude. As we showed in
\cite{xyzi}, the 2-mass-hard box rational parts can be expressed in
terms of 2-mass and 3-mass triangle integrals. If we write the
final results in terms of the 2-mass and 3-mass triangle
integrals, the analytic results are actually not too complicated and
would be well suited for inputs into any program for computing
physically interesting quantities.  One of the main goals of this paper
is to obtain explicit analytic results for the rational parts of
these two non-MHV helicity configurations. They are  the last
missing pieces for the complete partial helicity amplitudes of the
6-gluon one-loop QCD amplitude. Although our computation was done
by hand, we have also input our analytic results into Mathematica
codes. These codes would be useful for others to check our results
and as inputs to compute physically interesting quantities.

This paper is organized as follows: in the next section we recall
briefly the Feynman diagrams and the Feynman rules, tailored for
the purpose of computing the rational parts of the 6-gluon helicity
amplitude. A brief review of tensor reduction was given in
Sect.~3. The next 4 sections of the paper present the results with
some intermediate steps. In Sect.~8 we briefly discuss the
factorization properties of the one-loop amplitudes and the checks
done for the NMHV results. We collect some explicit results for
the rational parts of Feynman integrals in the Appendix.

\section{Notation, the Feynman diagrams and the Feynman rules}

A word about notation: we use the same notation as given in
\cite{xyzi}. We use $\epsilon_{i(i+1)\cdots (i+n)}$ to denote the
composite polarization vectors for sewing trees to the loop.

For the purpose of this paper we consider only the Feynman diagrams
and Feynman rules for the one-loop gluon amplitude with scalars
circulating in the loop. We do not follow the usual convention of
differentiating different particles by different kinds of lines
because there are only two kinds of particles: gluons and scalars,
and scalars only appear in the loop.

\begin{figure}[ht]
\centerline{\includegraphics[height=6cm]{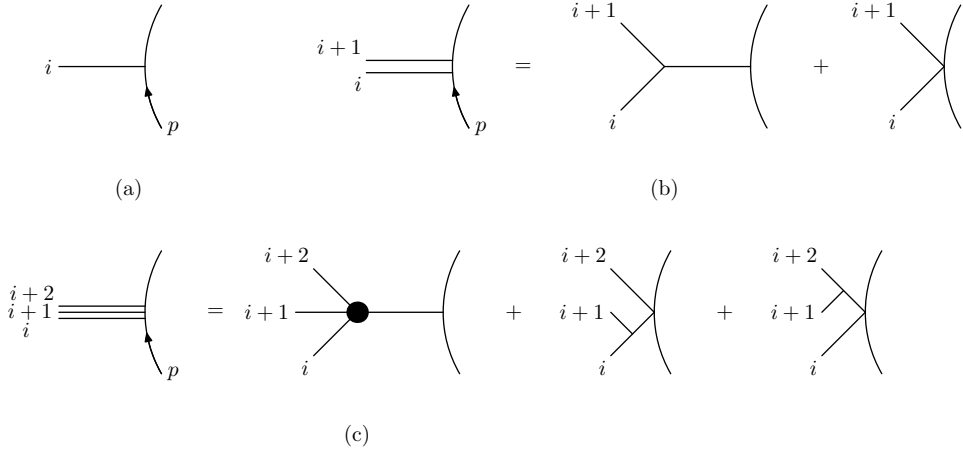} } \caption{The
Feynman rules for sewing trees to loop. The blob denotes an
expansion of tree amplitude.} \label{FeynmanTree}
\end{figure}

For explicit calculation of the one-loop amplitude by the usual
Feynman diagram technique, we can first collect all terms with the
same loop structure into one entity. Generally a few cyclicly
consecutive external lines are joined in tree diagrams and
connected to  the same point on the loop (sewing trees to loop).
We denote the sum of all these contributions by $P_{i(i+1)\cdots
(i+m-1)}$ for $m$ such external lines. For $m=1,2,3$, the relevant
Feynman diagrams are shown in Fig.~\ref{FeynmanTree}. Explicitly
we have:
\begin{eqnarray}
P_{i}(p) & = & (\epsilon_i,p) = (\epsilon_{i},p-k_i), \\
P_{i(i+1)}(p) & = & (\epsilon_{i(i+1)},p) - {1\over2} \,
 (\epsilon_i,\epsilon_{i+1})  , \\
P_{i(i+1)(i+2)}(p)
 & = & ( \epsilon_{i(i+1)(i+2)},p) - {1\over2}
\, (
    (\epsilon_{i(i+1)},\epsilon_{i+2}) +  (\epsilon_i,\epsilon_{(
    i+1)(i+2)}) ) ,
\end{eqnarray}
where $\epsilon_{\cdots}$'s are the composite polarization vectors
introduced in \cite{xyzi}. The computation of these composite
polarization vectors is a simplified version of the general
recursive calculation of the tree-level $n$-gluon amplitudes
\cite{BerendsGiele}.

\begin{figure}[ht]
\centerline{\includegraphics[height=8cm]{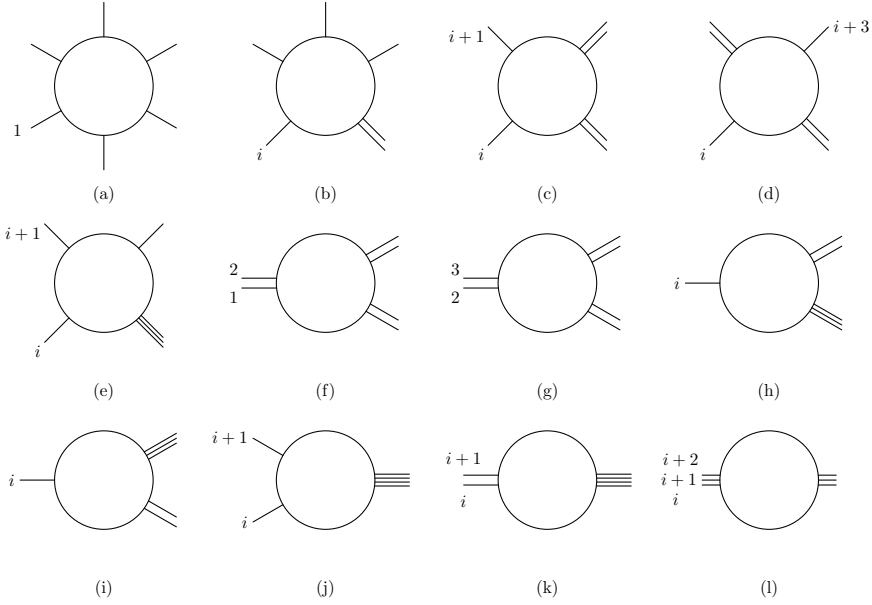} } \caption{All
the possible one-loop Feynman diagrams for six gluons. The index
$i$ can run from 1 to 3  in (d) and (l), and 1 to 6 in the rests
if there is an index $i$. } \label{FeynmanSixPoint}
\end{figure}

Considering all these different tree diagrams just as the same
diagram and denoting them by multiple parallel lines attached to
the loop, we have only 51 different Feynman diagrams for the
6-gluon one-loop amplitude (with scalars circulating in the loop).
Some representative diagrams are given in
Fig.~\ref{FeynmanSixPoint}. The counting goes as follows:
\begin{itemize}
\item 1 hexagon diagram, the diagram (a);
\item 6 pentagon diagrams because there are 6 different ways of combining
two consecutive external lines, diagram (b) with $i=1,\cdots,6$;
\item 15 box diagrams which are further divided into 6 two-mass-hard
box diagrams (diagram (c)), 3 two-mass-easy diagrams (diagram (d))
and 6 one-mass diagrams (diagram (e));
\item 20 triangle diagrams: 2 three-mass  triangle diagrams (diagrams (f) and (g)),
$6+6$ two-mass triangle diagrams (diagrams (h) and (i)) and 6 one-mass
triangle diagrams (diagram (j));
\item 9 bubble diagrams (diagrams (k) and (l)).
\end{itemize}

We use the notation $I_4^{2mh(i)}$, $I_3^{2m(i)}$ and $\tilde
I_3^{2m(i)}$ to denote the rational parts of the Feynman integrals
from the diagrams (c), (h) and (i). The rational parts from the
diagrams (f) and (g) are denoted by $\tilde I_3^{3m}$ and
$I_3^{3m}$ respectively. Explicit results from \cite{xyzi} are
collected in the Appendix for quick reference.

It is straightforward to compute the rational part from each
diagram. A better way to organize the computation is to use the
simple tensor reduction formulas derive in \cite{xyzi}. To simplify
things further we also organize the result into a summation over
the symmetry group. This is possible if the choice of the
reference momenta preserves the symmetry of the (rational
part of the) amplitude. The results for different Feynman diagrams
connected by symmetry operations can be obtained easily by using
symmetry operations on the rational parts directly. In this way we
need only to compute the different Feynman diagrams not connected
by symmetry operations. This roughly reduces the number of
different diagrams (needed to be computed explicitly) by a factor
of the rank of the (finite) symmetry group. Even for Feynman
diagrams which are invariant under the symmetry operations, it is
also possible to split the results into serval pieces which are
related by symmetry operations.

After we briefly review the tensor reduction formulas used in this
paper in the next section, we will then present some intermediate
steps and the results for the rational parts of the various (sets
of) Feynman diagrams for the 4 different helicity configurations
in 4 sections. At the beginning of each section we spell out the
symmetry group of the helicity configurations. The final result
for the rational part is written mostly as a summation over the
symmetry group. We give only the explicit result for the identity
group element. Others are obtained by symmetry actions. We will
indicate which diagram(s) gives the relevant contributions. We also
give the explicit choice of the polarization vectors for each
helicity configuration. Our choice of the polarization vectors
preserves the symmetry of the helicity configurations.

\section{Review of Tensor reduction of the one-loop amplitude}

There is a vast literature on this subject
\cite{PassarinoVeltman,Melrose:1965kb,BDKReduction,
BinothZ,Tarasov,Duplancic:2003tv,Denner}.  The tensor reduction relations we will
use for our calculation of the 5- and 6-gluon amplitudes are quite
simple. It is based on the BDK trick \cite{BDKX} of multiplying
and dividing by spinor square roots. We purposely made the
specific choice of the reference momenta in this paper to make all
tensor reductions simple enough to obtain relatively compact
analytic results for  (the rational parts of) the 6-gluon
amplitudes.

\begin{figure}[ht]
\centerline{\includegraphics[height=4cm]{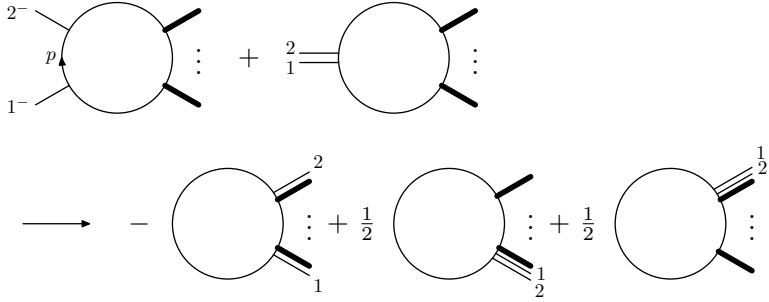} }
\caption{For two adjacent same helicities, the tensor reduction for
the combination of two diagrams is even simpler.} \label{Four}
\end{figure}

For tensor reductions with only 2 neighboring same helicity
external gluons, it is possible to choose the reference momenta to
be each other's momenta, i.e.  $\epsilon_1 =
\lambda_1\tilde\lambda_2$, $\epsilon_2 =
\lambda_2\tilde\lambda_1$. The tensor reduction is done by
considering the contributions from 2 diagrams together and the
result formula is shown pictorially in Fig.~\ref{Four}. The exact
algebraic formula is:
\begin{eqnarray}
 {(\epsilon_1,p+k_1)(\epsilon_2,p ) \over (p+k_1)^2
 p^2(p-k_{2 })^2} &  + & { (\epsilon_{12}, p+k_1) -
(\epsilon_1,\epsilon_2)/2 \over (p+k_1)^2 (p-k_{2})^2}
\nonumber \\
& = & - {1\over p^2} + { {1/ 2}\over  (p+k_1)^2} + {{1/2}\over
(p-k_{2})^2} . \label{eqreductiona}
\end{eqnarray}

\begin{figure}[ht]
\centerline{\includegraphics[height=3cm]{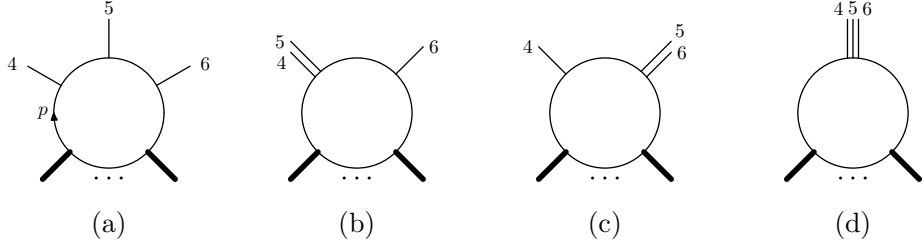} } \caption{For
three adjacent same helicities, the tensor reduction for the
combination of these four diagrams is also quite simple.}
\label{Five}
\end{figure}

For three adjacent particles with the same helicity, we choose the
following polarization vectors (omitting an overall factor for
each polarization vector):
\begin{equation}
\epsilon_4 = \lambda_5\tilde\lambda_4, \qquad \epsilon_5 =
\eta\tilde\lambda_5, \qquad \epsilon_6 = \lambda_5\tilde\lambda_6.
\end{equation}
Then the tensor reduction of the 4 diagrams shown in
Fig.~\ref{Five} is given as follows:
\begin{eqnarray}
A_{a} + A_{b} + A_{c} + A_{d} & = &
\frac{\langle\eta\,5\rangle}{2\, \langle6\,5\rangle}
\,{P_4(p) \over p^2\,(p-k_{4})^2}  \nonumber \\
& + & \frac{\langle\eta\,5\rangle}{2\, \langle4\,5\rangle}
\,{P_6(p-k_{45}) \over (p-k_{45})^2\,(p-k_{456})^2}  .
   \label{adjacentthree}
\end{eqnarray}
which has a nice symmetric property under the flipping operation $4
\leftrightarrow 6$.

Sometimes we also need the reduction formulas for $A_a+A_b$ or
$A_a+A_c$:
\begin{eqnarray}
A_{a} + A_{b} & =  & {P_4(p)\, P_5(p-k_4)\over
p^2\,(p-k_4)^2\,(p-k_{45})^2 } +
{P_{45}(p) \over p^2(p-k_{45})^2} \nonumber \\
& = &  - { (\eta\tilde\lambda_4,p)\over p^2\,(p-k_{4})^2} +
{\langle\eta\,5\rangle\over 2 \, \langle4\,5\rangle }\, \left[ {
1\over
(p-k_{45})^2} - {1\over p^2} \right]  ,  \\
A_{a} + A_{c} & =  & {P_5(\tilde p)\, P_6(\tilde p-k_5)\over
\tilde p^2\,(\tilde p-k_5)^2\,(\tilde p-k_{56})^2 } +
{P_{56}(\tilde p) \over \tilde p^2(\tilde p-k_{56})^2} \nonumber \\
& =&  { (\eta\tilde\lambda_6,\tilde p-k_5)\over (\tilde
p-k_5)^2\,(\tilde p-k_{56})^2} + {\langle\eta\,5\rangle\over 2\,
\langle6\,5\rangle }\, \left[ { 1\over
 \tilde p^2} - {1\over (\tilde p-k_{56})^2} \right],
 \end{eqnarray}
where $\tilde p = p - k_4$.

\section{MHV: $R(1^-2^+3^+4^-5^+6^+)$}
The reader is referred to eq. (31) in \cite{xyzi} for the precise
definition of the rational part, and it is the same to the next three sections.
For this helicity configuration, the symmetry group is
$$
G = \{ 1,\sigma,\tau,\sigma\tau\},
$$
where $\sigma = ( 2 \leftrightarrow 6, 3 \leftrightarrow 5)$ and
$\tau = ( i \to i+3)$ are the two generators of the symmetry
group. We have $\sigma\tau = ( 1 \leftrightarrow 4, 2
\leftrightarrow 3, 5 \leftrightarrow 6)$. Under the symmetry
action $\sigma$ (or $\sigma\tau$)  the ordering of the external
lines is reversed. The reversed ordering can be easily transformed
to the standard ordering (clockwise in our convention) by using
the symmetric or anti-symmetric property of the composite
polarization vectors. From this change of the ordering, there is
an overall sign of $(-)^n$ for an $n$-point amplitude which matches
the same factor from the reversing of the ordering of the color factor.

The polarization vectors for this helicity configuration can be
chosen as follows:
\begin{eqnarray}
\epsilon_2   =  {\lambda_3\, \tilde\lambda_2 \over \langle 3\,2
\rangle}, & & \epsilon_3   =  {\lambda_2\, \tilde\lambda_3 \over
\langle
2\,3 \rangle}, \\
\epsilon_5   =  {\lambda_6\tilde\lambda_5 \over \langle 6\,5
\rangle}, & & \epsilon_6   =  {\lambda_5\tilde\lambda_6 \over
\langle 5\,6 \rangle } .
\end{eqnarray}
We can leave the reference momenta of $\epsilon_{1,4}$ arbitrary. To be
specific, an
explicit symmetric choice is $\epsilon_{1} =
{\lambda_1\tilde\lambda_4\over[1\,4]}$ and $\epsilon_{4} =
{\lambda_4\tilde\lambda_1\over[4\,1]}$. To ease the writing,
we will omit all the denominators of these
polarization vectors and simply use $\epsilon_2=
\lambda_3\tilde\lambda_2$, etc. The overall factor will be
reinstated in the final result.

\begin{figure}[ht]
\centerline{\includegraphics[height=3.3cm]{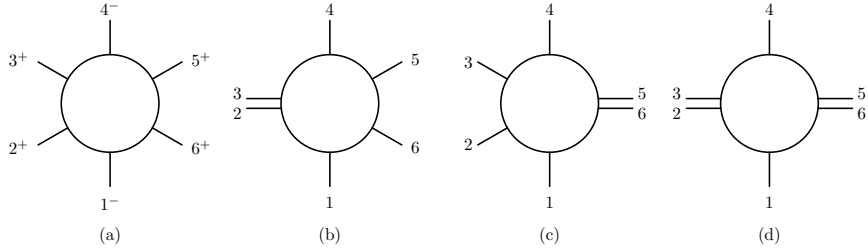} }
\caption{These 4 diagrams are tensor reduced simply with the two
pairs of external lines $k_{2,3}$ and $k_{5,6}$ with the same
helicities. } \label{xxe}
\end{figure}

\begin{figure}[ht]
\centerline{\includegraphics[height=5cm]{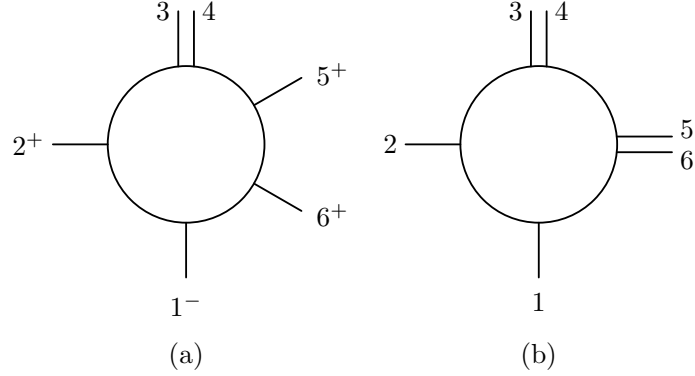} } \caption{This
set has 8 diagrams.  The other 6 diagrams are obtained by symmetry
operations. The above 2 diagrams can be reduced simply with
$k_{5,6}$.} \label{xxh}
\end{figure}

\begin{figure}[ht]
\centerline{\includegraphics[height=2.5cm]{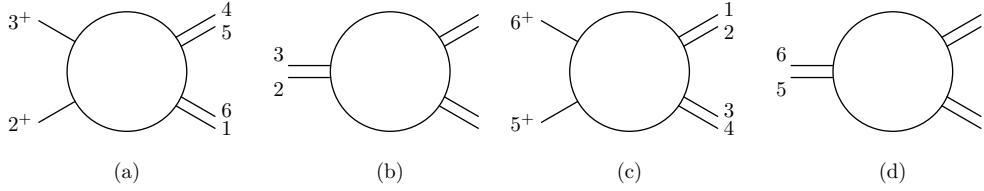} } \caption{This
set has 4 diagrams. The first 2 diagrams and the last 2 are
related by the symmetry operation $\tau$.} \label{xxg}
\end{figure}

\begin{figure}[ht]
\centerline{\includegraphics[height=3cm]{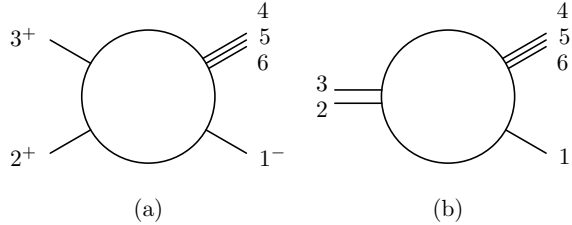} } \caption{This
set has 8 diagrams. The other 6 diagrams are obtained by symmetry
operations.} \label{xxf}
\end{figure}

\begin{figure}[ht]
\centerline{\includegraphics[height=2.5cm]{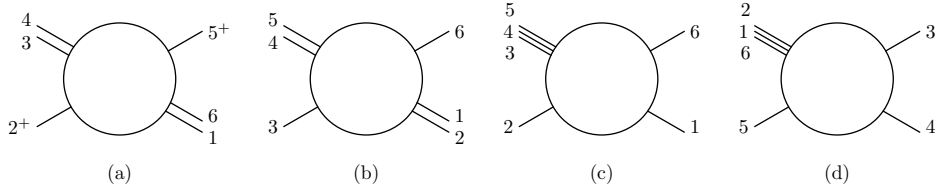} } \caption{This
set has 4 diagrams. (a) and (b) are related by symmetry operation
$\sigma$, whereas (c) and (d) are related by symmetry operation
$\tau$. } \label{xxi}
\end{figure}

\begin{figure}[ht]
\centerline{\includegraphics[height=3cm]{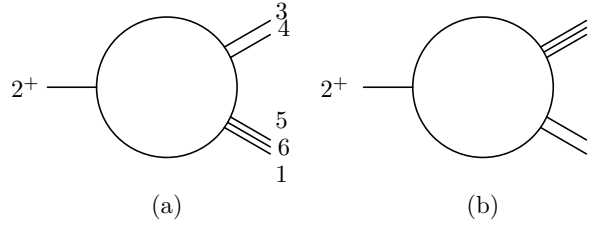} } \caption{This
set has 8 diagrams. Others are obtained from the above 2 diagrams
by symmetry operations. } \label{xxj}
\end{figure}

\begin{figure}[ht]
\centerline{\includegraphics[height=2.5cm]{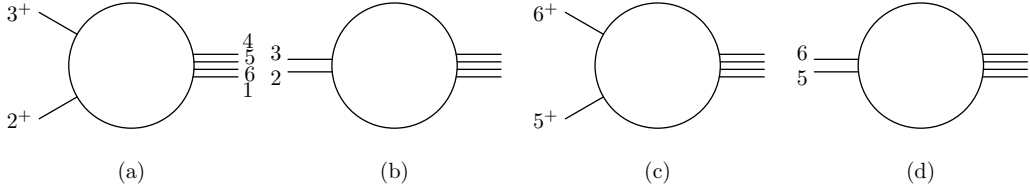} } \caption{This
set also has 4 diagrams. The last 2 are obtained from the first
two diagrams by the $\tau$ symmetry operation.} \label{xxxc}
\end{figure}

\begin{figure}[ht]
\centerline{\includegraphics[height=3cm]{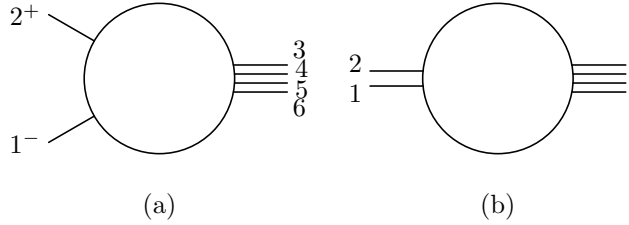} } \caption{This
set has 8 diagrams. The other 6 diagrams are obtained from these
two diagrams by symmetry operations. The rational part from
diagram (a) cancels the rational part from diagram (b).  }
\label{xxc}
\end{figure}

\begin{figure}[ht]
\centerline{\includegraphics[height=3cm]{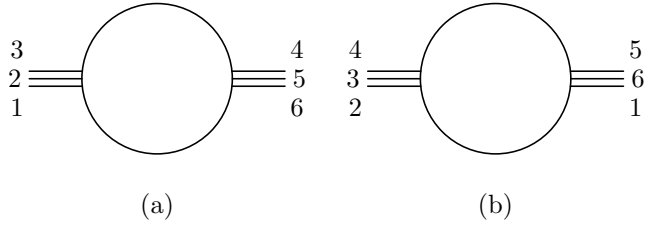} } \caption{This
set has 2 diagrams. The 2 diagrams are related by the $\sigma$
symmetry operation.} \label{xxb}
\end{figure}

\begin{figure}[ht]
\centerline{\includegraphics[height=3cm]{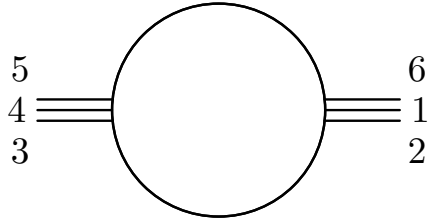} } \caption{This
diagram is the most symmetric one. It leads to a rational part
which is invariant under all symmetric operations.} \label{xxa}
\end{figure}

By using the above symmetry group $G$, the 51 Feynman diagrams are
classified into 10 sets. Representatives (or all Feynman diagrams)
from each set are shown
in Figs.~\ref{xxe} to \ref{xxa}. The diagrams are ordered from
higher point diagrams to lower point diagrams because the tensor
reduction of higher point diagrams often gives pieces which are
cancelled by lower point diagrams. Our strategy is to make this
(local) cancellation manifest by grouping these diagrams together.
We now compute the rational part of each set of diagrams in turn.

The 1st set consists of 4 Feynman diagrams as shown  in
Fig.~\ref{xxe}. The rational part can be  computed quite easily by
using the tensor reduction with the two pairs of external lines
$k_{2,3}$ and $k_{5,6}$ with the same helicities. The Feynman
diagrams reduce to only bubble diagrams. The rational part is:
\begin{eqnarray}
R_1 & = & - {1\over 72} \Big( 4 (\epsilon_1,k_{612}) \,
(\epsilon_4,k_{345}) + 2 \sum_{G}
(\epsilon_1,k_{12})\,(\epsilon_4,k_{12})
\nonumber \\
& +  & (\epsilon_1,k_{123})\, (\epsilon_4,k_{456}) +
(\epsilon_1,k_{561})\, (\epsilon_4,k_{234})  \Big) \nonumber \\
& + & {1 \over 36}\, (  2 (s_{12}  + s_{34} + s_{45}  + s_{61}  )
- 4 s_{345}  - s_{123} - s_{234}  ) \, (\epsilon_1,\epsilon_4).
\end{eqnarray}
The correction terms are cancelled between the 3 diagrams obtained
from the 1st step of the tensor reduction from hexagon (+ pentagon) to
box. By analyzing the symmetry property of the various terms, we
can write this result as a summation over the symmetry group G.
Explicitly we have
\begin{eqnarray}
R_1(1) & = & - {1\over 36} \left( (\epsilon_1,k_{2})\,
(\epsilon_4,k_{12}) + {1\over 2} (\epsilon_1,k_{612}) \,
(\epsilon_4,k_{345}) \right.
\nonumber \\
& + & \left . {1\over 4}
 (\epsilon_1,k_{123})\, (\epsilon_4,k_{456})  \right)
 +  {1\over 72}  ( 4 s_{12} - 2 s_{345} -   s_{123})
 \, (\epsilon_1,\epsilon_4) ,
\\
R_1(g)  & = &  g( R_1(1) ), \\
R_1  & = &  \sum_{g\in G} R_1(g).
\end{eqnarray}

The 2nd set consists of 8 diagrams and two representatives are
shown in Fig.~\ref{xxh}. The rational part from these two diagrams
gives $R_2(1)$ and we have:
\begin{eqnarray}
R_2(1) &  = &   {(\epsilon_3,\epsilon_4)(\epsilon_1,k_2) \over
8}\left( {2 (\epsilon_2,k_{61})  \over  s_{61} - s_{345}  } -
{(\epsilon_2,k_{34})  \over  s_{34} - s_{234}  }
+ {  (\epsilon_2,k_{1}) \over   s_{12} } \right) \nonumber \\
& -  & {1\over 18} ( (\epsilon_2, \epsilon_{34}) \epsilon_1 +
(\epsilon_2, \epsilon_{1}) \epsilon_{34} +
(\epsilon_1, \epsilon_{34}) \epsilon_2, k_5  - k_6)  \nonumber \\
&  - & {1\over24}\, ( (\epsilon_2,  \epsilon_{34})(\epsilon_1,k_2)
+
(\epsilon_2,  \epsilon_{1})(\epsilon_{34},k_2) ) \nonumber \\
&  + & {1\over 12  }\, (\epsilon_2, \epsilon_1)\, (
\epsilon_{34},k_1+2k_2)
\nonumber \\
& - & { s_{61} + s_{345} \over 6(s_{61} - s_{345})^2 } \,
(\epsilon_2, k_{61})
(\epsilon_{34},k_2)(\epsilon_1,k_2)  \nonumber \\
& - & { s_{34} + s_{234} \over 12(s_{34} - s_{234})^2 } \,
(\epsilon_2, k_{34})
(\epsilon_{34},k_2)(\epsilon_1,k_2)  \nonumber \\
& + & {(\epsilon_2, k_{61}) \over  6(s_{61} - s_{345}) } \, (
(\epsilon_{34},k_2) (\epsilon_1,k_{6}) -  (\epsilon_{34},k_{5})
(\epsilon_1,k_{2}) ) \nonumber \\
& - &  { (\epsilon_2, k_{34}) \over  12 (s_{34} - s_{234}) } \,
(\epsilon_{34},k_2) (\epsilon_1,k_{56})  \nonumber \\
& -  & { s_{61} + s_{345} \over 12(s_{61} - s_{345})  } \, (
(\epsilon_2,  \epsilon_{34})(\epsilon_1,k_2)   +
(\epsilon_2,  \epsilon_{1})(\epsilon_{34},k_2) ) \nonumber \\
& - & { s_{34} + s_{234} \over 24 (s_{34} - s_{234})  } \, (
(\epsilon_2,  \epsilon_{34})(\epsilon_1,k_2)  + (\epsilon_2,
\epsilon_{1})(\epsilon_{34},k_2) )  .
\end{eqnarray}

The computation of the rational part from the 4 diagrams shown in
Fig.~\ref{xxg} is also easy by using the tensor reduction with
$k_{2,3}$ or $k_{5,6}$. Here we need a correction term to the
naive tensor reduction in $D=4$ (see \cite{xyzi} for explicit
computation of the correction term). The result is:
\begin{eqnarray}
R_3(1) & = & R_3(\sigma\tau)  =
{1\over 36} \, (\epsilon_{45},k_3)\, (\epsilon_{61},k_2 ) \nonumber \\
& + &  {1\over 36} \, (2s_{345}  -s_{45}  - s_{61}  - 3 s_{23} )
\,
 (\epsilon_{45},\epsilon_{61})  .
\end{eqnarray}

The tensor reduction for the two diagrams shown in Fig.~\ref{xxf}
is also easy with the 2 external momenta $k_{2,3}$ because of the same
adjacent helicities. They are reduced  to bubble diagrams. As in
the last case with the reduction of the 2-mass-hard diagram, we
also need a correction term to the naive tensor reduction in
$D=4$. By including the correction term we have
\begin{equation}
R_4(1)  =
 {1\over 18} \, (\epsilon_{1},k_2)\, (\epsilon_{456},k_3 ) +
 {1\over 18} \, (2\,s_{12}  -s_{123}    - 3 s_{23} )
 \, (\epsilon_1,\epsilon_{456})   .
\end{equation}
In the above we have used the physical condition of the
(composite) polarization vectors. The term with the factor
$s_{23}$ comes from the correction term in the tensor reduction
from the (one-mass) box diagram. The other 6 diagrams (in 3 pairs)
give the contributions $R_4(\sigma)$, $R_4(\tau)$ and
$R_4(\sigma\tau)$.

The 4 diagrams shown in Fig.~\ref{xxi} give the most complicated
algebraic expressions for this helicity configuration. We give the
results in 2 parts:
\begin{eqnarray}
R_{5,1}(1)  & = &  { \langle 2\, 6 \rangle \over 4 \langle 2\, 5
\rangle } \left[ { (\epsilon_2, k_{34}) \over s_{34} - s_{234} } -
{ (\epsilon_2, k_{61}) \over s_{61} -s_{345} } \right]\,
(\epsilon_3,\epsilon_4)(\epsilon_{61},k_2)
\nonumber \\
& + & { \langle 2\, 6 \rangle \over 4  \langle 2\, 5 \rangle }
\left[ { (\epsilon_2, k_{34}) \over s_{34} - s_{234} } - {
(\epsilon_2, k_{61}) \over s_{61} -s_{345} } \right]\,
(\epsilon_1,\epsilon_6)(\epsilon_{34},k_2)
\nonumber \\
& + &  { \langle 2\, 5 \rangle \over 4  \langle 2\, 6 \rangle }
\left[ {  (\epsilon_2, k_{61}) \over  s_{61} - s_{345}  } +
 {  (\epsilon_2, k_{1}) \over s_{12}   }  \right] (\epsilon_{1},k_2)
( (\epsilon_3,\epsilon_{45})  +
(\epsilon_{34},\epsilon_{5}) ) \nonumber \\
& + &  { \langle 2\, 6 \rangle \over 9  \langle 2\, 5 \rangle }
\big(
 (\epsilon_2,\epsilon_{34}) \epsilon_{61}  +
(\epsilon_2,\epsilon_{61}) \epsilon_{34}  +
 (\epsilon_{34},\epsilon_{61}) \epsilon_{2}, k_5\big)
\nonumber \\
& + &  { \langle 2\, 5 \rangle \over    \langle 2\, 6 \rangle }
\left[ {1\over 9}\,  \big(
 (\epsilon_2,\epsilon_{345}) \epsilon_{1}  +
(\epsilon_2,\epsilon_{1}) \epsilon_{345}  +
 (\epsilon_{345},\epsilon_{1}) \epsilon_{2}, k_6\big) \right. \nonumber \\
& - & {1\over 12} \, \big(
 (\epsilon_2,\epsilon_{345}) \epsilon_{1}  +
(\epsilon_2,\epsilon_{1}) \epsilon_{345}, k_2\big)   \left. +
{1\over 6  } \, (\epsilon_1, \epsilon_2 )\,
(\epsilon_{345},k_2-k_6) \right]
\nonumber \\
& + & { \langle 2\, 6 \rangle \over    \langle 2\, 5 \rangle }
\left[   {1\over 12} \, \left[ {s_{34} + s_{234} \over s_{34} -
s_{234} } + {s_{61} +s_{345}\over s_{61} - s_{345}} \right]
((\epsilon_2,\epsilon_{34}) \epsilon_{61} +
 (\epsilon_2,\epsilon_{61}) \epsilon_{34}, k_2) \right.
 \nonumber \\
 & +  & (\epsilon_{34},k_2)(\epsilon_{61},k_2) \, \left[
{(s_{34} + s_{234})(\epsilon_{2},k_{34})\over 6(s_{34} -
s_{234})^2 } + { (s_{61} + s_{345})(\epsilon_{2},k_{61})\over
6(s_{61} - s_{345})^2 } \right]
\nonumber \\
& + & \left.   \left[ { (\epsilon_{2},k_{34})
(\epsilon_{34},k_2)(\epsilon_{61},k_5)
 \over 6(s_{34} - s_{234})  }
+{  (\epsilon_{2},k_{61}) (\epsilon_{34},k_5)(\epsilon_{61},k_2)
\over 6(s_{61} - s_{345})  }
\right] \right] \nonumber \\
& - & { \langle 2\, 5 \rangle \over    \langle 2\, 6 \rangle }
\left[ {s_{61} +s_{345}\over 12( s_{61} - s_{345})}
((\epsilon_2,\epsilon_{345}) \epsilon_{1} +
 (\epsilon_2,\epsilon_{1}) \epsilon_{345}, k_2) \right.
 \nonumber \\
 & +  & \left. (\epsilon_2,k_{61})(\epsilon_{345},k_2)  \left[
 { (s_{61} + s_{345})(\epsilon_{1},k_{2})\over 6(s_{61} - s_{345})^2
 }-
 { (\epsilon_{1},k_6)  \over 6(s_{61} - s_{345})  }
\right] \right] ,
\end{eqnarray}
and
\begin{eqnarray}
R_{5,2}(1)   & = & (k_2, k_5) \, (\epsilon_{34},\epsilon_{61}) \,
\left[ { 5\over 18} \,
 { \langle2\, 3 \rangle  \langle5\, 6\rangle  \over   \langle 2\, 5 \rangle^2}-
  {1\over 18} \, {\langle3\, 5 \rangle  \langle2\, 6\rangle \over
  \langle 2\, 5 \rangle^2}
  \right]  \nonumber \\
 & +  &  \big( (\epsilon_{34},k_2 )(\epsilon_{61}, k_5) +
 (\epsilon_{34},k_5)(\epsilon_{61}, k_2)\big) \nonumber \\
 & & \times  \,\left[ -
{ 2\over 9} \,
 { \langle2\, 3 \rangle  \langle5\, 6\rangle  \over   \langle 2\, 5 \rangle^2} +
  {1\over 36} \, {\langle3\, 5 \rangle  \langle2\, 6\rangle \over
   \langle 2\, 5 \rangle^2}
  \right]  \nonumber \\
& + &  (k_2,k_6)(\epsilon_{1},\epsilon_{345}) \, \left[ - { 5\over
18} \,
 { \langle2\, 3 \rangle  \langle5\, 6\rangle  \over   \langle 2\, 6 \rangle^2}-
  {1\over 18} \, {\langle3\, 6 \rangle  \langle2\, 5\rangle \over
  \langle 2\, 6 \rangle^2}
  \right]
  \nonumber \\
& +  &  (\epsilon_{1},k_2 )(\epsilon_{345}, k_6) \,\left[
  { 4\over 9} \,
 { \langle2\, 3 \rangle  \langle5\, 6\rangle  \over   \langle 2\, 6 \rangle^2} +
  {1\over 18} \, {\langle3\, 6 \rangle  \langle2\, 5\rangle \over
   \langle 2\, 6 \rangle^2}
  \right]  \nonumber \\
 & + & (\epsilon_1,k_2)(\epsilon_{345},k_2)  \,
 { \langle2\, 3 \rangle \langle 5\, 6\rangle
  \over 4 \, \langle 2\, 6\rangle^2 } + {s_{61} + s_{345} \over s_{61}  -
 s_{345} } \nonumber \\
 & & \times \left[  { \langle2\, 3 \rangle \langle 5\, 6\rangle
 \over 4\, \langle2\, 5 \rangle^2}  \,  (\epsilon_{34}, k_2)(\epsilon_{61}, k_2)
  +   { \langle2\, 3 \rangle \langle 5\, 6\rangle
 \over 4\, \langle2\, 6 \rangle^2}  \, (\epsilon_{345}, k_2)(\epsilon_{1}, k_2)
  \right] \nonumber \\
  & + & { s_{34} + s_{234} \over   s_{34} -  s_{234} }
  \,  { \langle2\, 3 \rangle \langle 5\, 6\rangle
 \over  4\, \langle 2\, 5\rangle^2 } \, (\epsilon_{34},k_2)(\epsilon_{61},k_2) .
\end{eqnarray}
The complete rational part is $R_5(1) = R_{5,1}(1) + R_{5,2}(1)$.

The rational part from the 2-mass triangle diagrams shown in
Fig.~\ref{xxj} can be written directly by using
eqs.~(\ref{twomasstwo}) and (\ref{twomassthree}). We have
\begin{eqnarray}
R_{6,1}(1) &  = &  {1\over 36} \, ( 7 (\epsilon_2, \epsilon_{34})
(\epsilon_{561}, k_{2}) - 7 (\epsilon_2, \epsilon_{561})
(\epsilon_{34}, k_{2}) + 4 (   \epsilon_{34},
\epsilon_{561})(\epsilon_{2}, k_{34}) )
\nonumber \\
& - &  {s_{34} +  s_{234} \over 12(s_{34} -  s_{234})  } \, (
(\epsilon_2,  \epsilon_{34}) (\epsilon_{561}, k_{2}) +
(\epsilon_2,  \epsilon_{561}) (\epsilon_{34}, k_{2}) )
\nonumber \\
& - &    {s_{34} +  s_{234} \over 6(s_{34} -  s_{234})^2 } \,
(\epsilon_2, k_{34})(\epsilon_{34}, k_{2})(\epsilon_{561}, k_{2})
\nonumber \\
&  - & {1\over 4} (\epsilon_3,\epsilon_4)\left[
(\epsilon_2,\epsilon_{561}) + {(\epsilon_2,k_{34})(\epsilon_{561}
, k_2) \over s_{34} - s_{234} } \right]
\nonumber \\
&  - & {1\over 4} ( (\epsilon_5,\epsilon_{61} )+
(\epsilon_{56},\epsilon_{1}) ) \left[ (\epsilon_2,\epsilon_{34}) +
{(\epsilon_2,k_{34})(\epsilon_{34} , k_2) \over s_{34} - s_{234} }
\right] ,
\end{eqnarray}

\begin{eqnarray}
R_{6,2}(1)  &  = &  {1\over 36} \, ( 7 (\epsilon_2,
\epsilon_{345})
 (\epsilon_{61}, k_{2}) -
7 (\epsilon_2,  \epsilon_{61}) (\epsilon_{345}, k_{2}) - 4 (
\epsilon_{345}, \epsilon_{61})(\epsilon_{2}, k_{61}) )
\nonumber \\
& + &  {  s_{61} + s_{345}  \over 12( s_{61} - s_{345} )  } \, (
(\epsilon_2,  \epsilon_{345}) (\epsilon_{61}, k_{2}) +
(\epsilon_2,  \epsilon_{61}) (\epsilon_{345}, k_{2}) )
\nonumber \\
& + &    { s_{61} + s_{345} \over 6( s_{61} -s_{345} )^2 } \,
(\epsilon_2, k_{61})(\epsilon_{345}, k_{2})(\epsilon_{61}, k_{2})
\nonumber \\
&  - & {1\over 4}
(\epsilon_6,\epsilon_1)\left[(\epsilon_2,\epsilon_{345}) +
{(\epsilon_2,k_{61})(\epsilon_{345} , k_2) \over s_{61} -  s_{345}
} \right]
\nonumber \\
&  - & {1\over 4} ( (\epsilon_3,\epsilon_{45} )+
(\epsilon_{34},\epsilon_{5}) ) \left[ (\epsilon_2,\epsilon_{61}) +
{(\epsilon_2,k_{61})(\epsilon_{61} , k_2) \over  s_{61} - s_{345}
} \right]  , \\
R_6(1) & = & R_{6,1}(1) +  R_{6,2}(1).
\end{eqnarray}

The diagrams in Fig.~\ref{xxxc} can be easily computed either directly or by using the
tensor reduction formula with correction terms. We have:
\begin{eqnarray}
R_7(1) = R_7(\sigma\tau) & = &
 {1\over 8}\, s_{23}\, ( (\epsilon_{45}, \epsilon_{61})
 + (\epsilon_4,\epsilon_{561}) +
 (\epsilon_{456},\epsilon_1) )
 \nonumber  \\
 &  -  & {1 \over 12} \, s_{23} (\epsilon_{4561}, k_2)  .
\end{eqnarray}

The rational parts from the two diagrams in Fig.~\ref{xxc} have
a very nice property. By using the relation
$(\epsilon_1,k_2)\,(\epsilon_2,k_1)=s_{12}(\epsilon_1,\epsilon_2)$
for a pair of  neighboring lines with opposite helicities, one
shows that they cancel each other and the sum of the rational
parts is identically equal to zero, i.e.,
\begin{equation}
R_8(1) =0. \label{eqzero}
\end{equation}
This result
is also true for the other helicity configurations as long as the
2 massless external lines in the 1-mass triangle diagram have
opposite helicities.

For the diagrams in Fig.~\ref{xxb}, we have
\begin{eqnarray}
R_9(1) & = & R_9(\tau) = - {1\over 18} \, s_{123} (\epsilon_{123}, \epsilon_{456}), \\
R_9(\sigma) & = & R_9(\sigma\tau) = R_9(1)|_{2\leftrightarrow6,
3\leftrightarrow5} = -{1\over18} \, s_{234}\, (\epsilon_{234},
\epsilon_{561}) .
\end{eqnarray}
In the above, the indicated permutations are applied to the
polarization vectors, momenta and kinematic variables.

The diagram shown in Fig.~\ref{xxa}  is invariant under all
symmetry operations. The rational part is:
\begin{equation}
R_{10}   =  - {1\over 9}\, s_{345}(\epsilon_{345},
\epsilon_{612}).
\end{equation}
Dividing by the rank of the (finite) symmetry group gives $R_{10}(g)$:
\begin{equation}
R_{10}(g)   =   - {1\over 36}\, s_{345}(\epsilon_{345},
\epsilon_{612}) , \qquad g =  1,\sigma,\tau,~{\rm or}~\sigma\tau .
\end{equation}

Combining all the above results together, the complete rational
part is given as follows:
\begin{equation}
R = F\, \sum_{g\in G} R(g) = F\, \sum_{g\in G} \sum_{i=1}^{10} R_i(g),
\end{equation}
where
\begin{equation}
F = -{1\over (\langle2\,3\rangle\,\langle5\,6\rangle\,[1\,4])^2},
\end{equation}
is an overall factor from the omitted factors of the polarization
vectors. The explicit form of $R$ is not quite illuminating and we
will not write it down here. To be sure that the above result is
correct we have chosen another completely different set of
reference momenta for the polarization vectors and obtained the
same analytic result (in a quite different form of course). To do
some consistency checks, let us consider the double pole terms
from $R(1)$:
\begin{eqnarray}
A_2(1) & = &  { s_{61} + s_{345} \over 6( s_{61} -s_{345} )^2 }
\Big[ (\epsilon_2, k_{61})(\epsilon_{345}, k_{2})(\epsilon_{61},
k_{2})
  -  (\epsilon_2, k_{61}) (\epsilon_{34},k_2)(\epsilon_1,k_2)
 \nonumber \\
 &  -  &   { \langle 2\, 5 \rangle \over    \langle 2\, 6 \rangle }
  (\epsilon_2,k_{61})(\epsilon_{345},k_2) (\epsilon_{1},k_2)
  + { \langle 2\, 6 \rangle \over    \langle 2\, 5 \rangle }  \,
(\epsilon_{34},k_2)(\epsilon_{61},k_2)  (\epsilon_{2},k_{61}) \Big] \nonumber \\
& + & {(s_{34} + s_{234}) \over 6(s_{34} - s_{234})^2 } \left[ {
\langle 2\, 6 \rangle \over    \langle 2\, 5 \rangle }  \,
(\epsilon_{34},k_2)(\epsilon_{61},k_2) \,
 (\epsilon_{2},k_{34}) \right. \nonumber \\
 & - &  \left.  {  1 \over  2 } \, (\epsilon_2, k_{34})
(\epsilon_{34},k_2)(\epsilon_1,k_2)  -
 (\epsilon_2, k_{34})(\epsilon_{34}, k_{2})(\epsilon_{561},
 k_{2})\right].
\end{eqnarray}
We have checked that these terms agree with the result of the
cut-constructible part  \cite{BBST} ($T_3$ terms) as required by
the absence of spurious poles in the complete amplitude
(cut-constructible part $+$ rational part), up to simple pole
terms. These residual simple pole terms added together with the
simple pole terms from our explicit computations also agree with
the cut-constructible part  of \cite{BBST} ($T_2$ terms).

In comparing with the notation of Bern, Dixon and Kosower
\cite{BDKA}, $R$ is one half of what they called $R(z)$. The exact
relation is:
\begin{equation}
R = {1\over 2} \, \left(  R(z)|_{z=0} + {2\over 9}\, A^{\rm tree}
\right),
\end{equation}
by striping away a factor of $i\,c_{\Gamma}$.  By using the
Mathematica code provided by them for their result, we have done a
numerical check by comparing our result with the result of Berger,
Bern, Dixon, Forde and Kosower  \cite{BDKE}. The check is done by
randomly assigning a set of complex rational numbers to the 12
independent $\langle i\,j\rangle$ and $[i\,j]$. We found exact
match without suffering the usual real number approximation
because all computations are done with rational numbers.

We note that this numerical check actually gives a proof of the
equivalence of  our result and  the result of \cite{BDKE}. This is
because a rational function with a limited number of different
denominators and a fixed degree can only depends on a finite
number of free parameters. Although we have no exact estimation of
the number of these free parameters, the number (over a few
thousands) of checks we have done should be quite enough.

\section{MHV: $R(1^-2^+3^-4^+5^+6^+)$}
For this helicity configuration, the symmetry group is
$$
G = \{ 1, \sigma \},
$$
where $\sigma = ( 1 \leftrightarrow 3, \, 4\leftrightarrow 6 )$.
The polarization vectors are:
\begin{eqnarray}
\epsilon_1   =  {\lambda_1\,\tilde\lambda_2 \over [1\,2] }, & &
\epsilon_3   =   {\lambda_3\,\tilde\lambda_2 \over [3\,2] }, \\
\epsilon_2   =  {\lambda_5\, \tilde\lambda_2 \over \langle 5\,2
\rangle }, & & \epsilon_5   =  {\lambda_2\, \tilde\lambda_5 \over
\langle 2\,5 \rangle }, \\
\epsilon_4   =  {\lambda_5\, \tilde\lambda_4 \over\langle 5\,4
\rangle }, & & \epsilon_6   =  {\lambda_5\, \tilde\lambda_6 \over
\langle 5\,6 \rangle } .
\end{eqnarray}
We note that the above choice of reference momenta preserves the symmetry
of the amplitude manifestly.

For this helicity configuration,
there are some diagrams which are $G$-invariant. Even for these diagrams
it still simplifies to split  the result into 2 parts which are
related by the symmetry operation $\sigma$. In the following we will present the
results of the rational parts from the various diagrams in a uniform way such that the
complete result is obtained by summing over various individual terms
and over the symmetry group, although the symmetry group $G$ only has rank 2.

The 51 Feynman diagrams are split into 12 sets. We now give the
results for the rational part from each set in turn, with some
explanations of the tricks we used.

\begin{figure}[ht]
\centerline{\includegraphics[height=3.5cm]{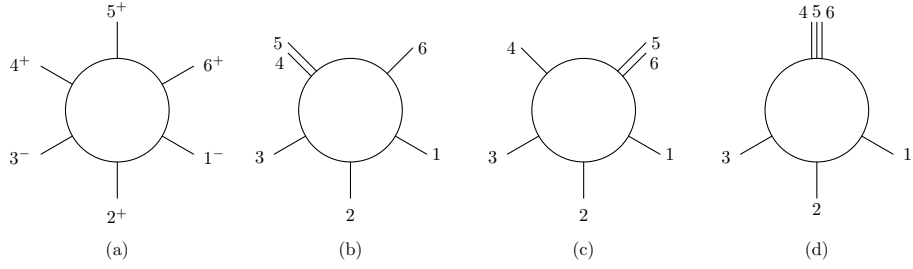} }
\caption{This set has 4 Feynman diagrams. See text for the logic
of this combination.} \label{NMHVaa}
\end{figure}

The first set shown in Fig.~\ref{NMHVaa} has 4 Feynman diagrams.
This includes the sole 6-point diagram. The logic for considering these four diagrams together is
that the tensor reduction gives only two one-mass box diagrams which can be easily
computed following the method elucidated in \cite{xyzii}. The rational part is:
\begin{eqnarray}
R_1(1) &=&\frac {\la25\ra^2\la56\ra}{8\la26\ra^2\la45\ra}\left[
1+{s_{12}+s_{345} \over s_{12}-s_{345} }\right] \,
(\epsilon_1,k_6)(\epsilon_3,k_6)
\nonumber\\
&+ &  \frac {\la25\ra\la56\ra}{6\la26\ra\la45\ra}\left(
{(\epsilon_1,k_6)\over s_{12}-s_{345}
}(\epsilon_3,k_{45})(\epsilon_6,k_{12}) \right. \nonumber\\&& -\frac 1 2
{s_{12}+s_{345} \over (s_{12}-s_{345})^2
}(\epsilon_3,k_6)(\epsilon_6,k_{12})(\epsilon_1,k_{6})  \nonumber\\
&& -\frac 1 4 {s_{12}+s_{345} \over s_{12}-s_{345}
}((\epsilon_3,k_6)(\epsilon_6,\epsilon_1)+(\epsilon_1,k_6)(\epsilon_6,\epsilon_3))
\nonumber\\&& - \frac 1
2(\epsilon_1,\epsilon_6)(\epsilon_3,2k_1+k_6) \nonumber \\&&
-  \left. \frac 1 4
((\epsilon_1,\epsilon_6)(\epsilon_3,k_6)+(\epsilon_3,\epsilon_6)(\epsilon_1,k_6))\right)
\end{eqnarray}

\begin{figure}[ht]
\centerline{\includegraphics[height=3cm]{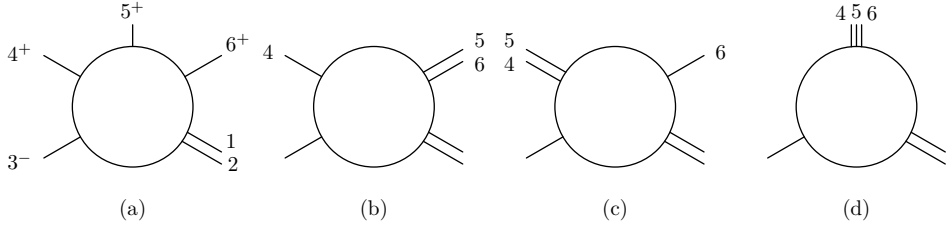} }
\caption{This set has 8 diagrams. The other 4 are obtained by the
symmetry operation $\sigma$.} \label{NMHVac}
\end{figure}

\begin{figure}[ht]
\centerline{\includegraphics[height=2.7cm]{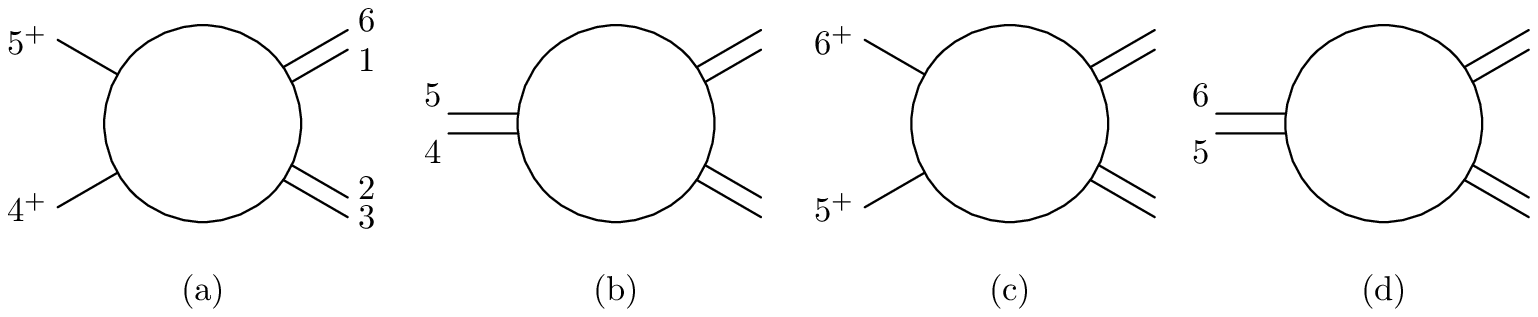} }
\caption{This set has 4 diagrams. The first 2 and the last 2 are
related by symmetry operation $\sigma$.} \label{NMHVae}
\end{figure}

The 4 diagrams shown in Fig.~\ref{NMHVac} can be reduced similarly as in the above
and the rational part is:
\begin{eqnarray}
R_2(1) & = & \frac {\la25\ra}{6\la45\ra}\left({(\epsilon_3,k_{45})\over
s_{12}-s_{345}}(\epsilon_{6},k_{12})(\epsilon_{12},k_6) \right . \nonumber
\\&& -\frac 1 2{s_{12}+s_{345}\over
(s_{12}-s_{345})^2}(\epsilon_3,k_{6})(\epsilon_{6},k_{12})(\epsilon_{12},k_6)
\nonumber \\ && -\frac 1 4{s_{12}+s_{345}\over
s_{12}-s_{345}}((\epsilon_{6},\epsilon_{12})(\epsilon_3,k_{6})
+(\epsilon_{6},\epsilon_{3})(\epsilon_{12},k_{6})) \nonumber\\ &&
- \left. \frac 7
{12}((\epsilon_{6},\epsilon_{12})(\epsilon_{3},2k_1+k_{6})
+(\epsilon_{6},\epsilon_{3})(\epsilon_{12},k_{6}))\right) .
\end{eqnarray}
by omitting a one-mass triangle diagram which will be cancelled
by a bubble diagram from the tensor reduction of Fig.~\ref{NMHVae}.

\begin{figure}[ht]
\centerline{\includegraphics[height=2.7cm]{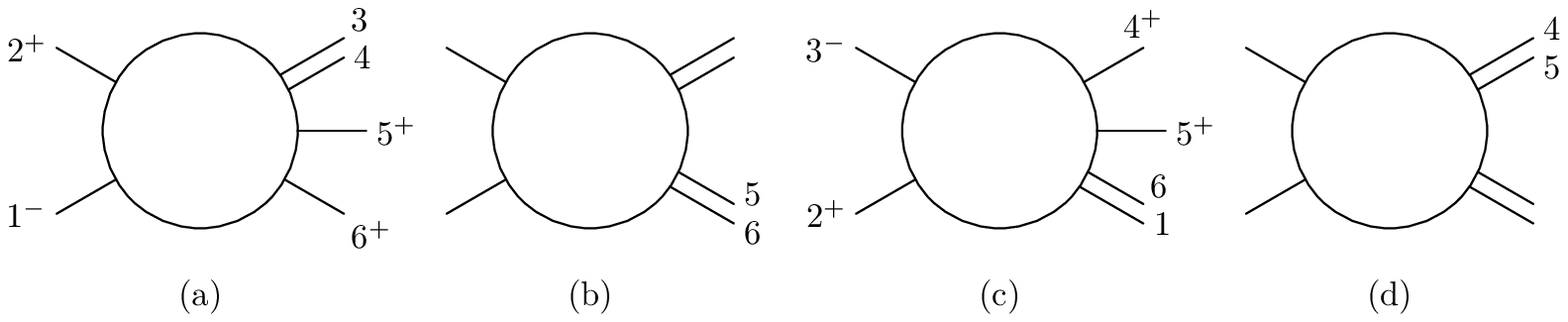} }
\caption{This set has 4 diagrams. The first 2 and the last 2 are
related by symmetry operation $\sigma$.} \label{NMHVaxx}
\end{figure}

The rational part from the diagrams in Fig.~\ref{NMHVaxx} is:
\begin{eqnarray}
R_{3}(1) &=&-\frac {\la25\ra}{\la26\ra}\left[ -\frac 1
4\left(1+{s_{12}+s_{345}\over
s_{12}-s_{345}}\right)(\epsilon_{1},k_{6})(\epsilon_{34},k_{6})
\right. \nonumber \\&& + \left. \frac 1 9
(5(k_{2},k_{6})(\epsilon_{34},\epsilon_{1})-4(\epsilon_{1},k_{6})(\epsilon_{34},k_2))
\right]  \nonumber \\&+&\frac {\la56\ra}{3\la26\ra}\left[
 {(\epsilon_{34},k_{5})\over s_{12}-s_{345}}(\epsilon_{1},k_{6})(\tilde{\epsilon}_{6},k_{1})
\right. \nonumber\\&& -\frac 1 2{s_{12}+s_{345}\over
(s_{12}-s_{345})^2}(\epsilon_{1},k_{6})(\epsilon_{34},k_{6})(\tilde{\epsilon}_{6},k_{1})
\nonumber \\&& -\frac 1 {4}\left({s_{12}+s_{345}\over
s_{12}-s_{345}}+1\right)((\epsilon_{34},k_{6})(\tilde{\epsilon}_{6},\epsilon_{1})
+(\epsilon_{1},k_{6})(\tilde{\epsilon}_{6},\epsilon_{34}))
\nonumber\\ &&
 -\frac 1 2(\tilde{\epsilon}_{6},\epsilon_{1})(\epsilon_{34},k_1-k_5)
- \left. \frac2
3(\tilde{\epsilon}_{6},\epsilon_{1})(\epsilon_{34},k_{2}) \right]
\nonumber\\&  - & \frac
{\la56\ra}{4\la26\ra}(\epsilon_{3},\epsilon_{4})(\tilde{\epsilon}_{6},\epsilon_{1})\left(1
+{s_{61}\over s_{12}-s_{345}}\right) \nonumber \\
& - & \frac {\la25\ra}{12\la45\ra}\left[ \frac2
3((\epsilon_{2},\epsilon_{61})(\epsilon_{3},k_{61})
+(\epsilon_{2},k_{61})(\epsilon_{3},\epsilon_{61})) \right.
\nonumber\\&& + \left. {(\epsilon_{3},k_{61})\over
s_{61}-s_{345}}(\epsilon_{2},k_{61})(\epsilon_{61},k_{2})\right] ,
\end{eqnarray}
by also omitting a one-mass triangle diagram which will be cancelled
by a bubble diagram from the tensor reduction of Fig.~\ref{NMHVae}. Here
$\tilde\epsilon_6 = \lambda_2\tilde\lambda_6$.

\begin{figure}[ht]
\centerline{\includegraphics[height=3cm]{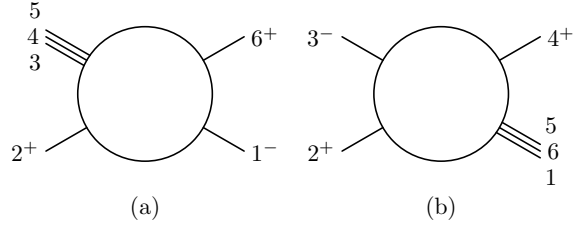} }
\caption{This set has 2 diagrams which are related by the symmetry
operation $\sigma$.} \label{NMHVab}
\end{figure}

The 4th set  shown in Fig.~\ref{NMHVab} has 2 diagrams which
are related by the symmetry operation $\sigma$. The rational part
from the first diagram is:
\begin{eqnarray}
R_4(1) & = &-\frac {\la56\ra}{4\la26\ra}\left[
{(\epsilon_1,k_6)(\epsilon_6,k_{12})\over s_{12}-s_{345}
}+(\epsilon_1,\epsilon_6)\right]
((\epsilon_3,\epsilon_{45})+(\epsilon_{34},\epsilon_5))
\nonumber\\
& - & \frac {\la25\ra\la56\ra}{4\la26\ra^2}\left[
-{s_{12}+s_{345} \over s_{12}-s_{345}
}(\epsilon_1,k_6)(\epsilon_{345},k_{6}) \right. \nonumber\\&&
\left. +\frac 8 3 (k_2,k_6)(\epsilon_1,\epsilon_{345})-
(\epsilon_1,k_6)(\epsilon_{345},2k_{2}+k_{6})\right] \nonumber
\\&& +\frac {\la56\ra}{3\la26\ra}\left[ -\frac 1 2{s_{12}+s_{345}
\over (s_{12}-s_{345})^2
}(\epsilon_1,k_{6})(\epsilon_6,k_{12})(\epsilon_{345},k_{6})
\right. \nonumber \\&& -\frac 1 4\left( {s_{12}+s_{345} \over
 s_{12}-s_{345}
}+1\right)((\epsilon_6,\epsilon_{345})(\epsilon_1,k_{6})
+(\epsilon_6,\epsilon_{1})(\epsilon_{345},k_{6})) \nonumber \\&&
-\left. \frac 1 2(\epsilon_6,\epsilon_{1})(\epsilon_{345},k_1+2k_2) +
\frac 1
3((\epsilon_1,\epsilon_{345})(\epsilon_6,k_{2})+(\epsilon_6,\epsilon_{1})(\epsilon_{345},k_2))
\right]  \nonumber \\ && +\frac {\la25\ra}{3\la26\ra}\left[ \frac 1
3((\epsilon_1,\epsilon_{345})(\epsilon_2,k_{6})
+(\epsilon_2,\epsilon_{345})(\epsilon_1,k_{6})) \right. \nonumber \\ &&
+ \left. \frac 1 2 {(\epsilon_{345},k_2)\over (s_{61}-s_{345})
}(\epsilon_1,k_{6})(\epsilon_2,k_{61}) \right]  .
\end{eqnarray}

The rational part from the diagrams in Fig.~\ref{NMHVae} is:
\begin{eqnarray}
R_5(1) &=& -\frac1 6\la25\ra[56](\epsilon_{12},\epsilon_{34}) -
 {(\tilde{\epsilon}_{6},k_{1})(\epsilon_{12},k_{6})\over
12(s_{12}-s_{345})}(3(\epsilon_{3},\epsilon_{4})
-4(\epsilon_{34},k_{5})) \nonumber\\&- & \frac1
6{s_{12}+s_{345}\over
(s_{12}-s_{345})^2}(\tilde{\epsilon}_{6},k_{1})(\epsilon_{12},k_{6})(\epsilon_{34},k_{6})
\nonumber\\&- & \frac1 {12}{s_{12}+s_{345}\over
s_{12}-s_{345}}((\tilde{\epsilon}_{6},\epsilon_{34})(\epsilon_{12},k_{6})
+(\tilde{\epsilon}_{6},\epsilon_{12})(\epsilon_{34},k_{6}))
\nonumber\\&+ & \frac
1{36}(-7(\tilde{\epsilon}_{6},\epsilon_{34})(\epsilon_{12},k_{6})
+7(\tilde{\epsilon}_{6},\epsilon_{12})(\epsilon_{34},2k_5+k_{6})
\nonumber \\ &  & + 4(\tilde{\epsilon}_{6},k_1)(\epsilon_{12},\epsilon_{34})
-9(\tilde{\epsilon}_{6},\epsilon_{12})(\epsilon_{3},\epsilon_{4})
)  ,
\end{eqnarray}
where $\tilde\epsilon_6 = \lambda_2\tilde\lambda_6$ as defined before. As we said before,
the bubble diagrams from tensor reduction cancel the rational part from the one-mass triangle
diagrams from tensor reduction in Figs.~\ref{NMHVac} and \ref{NMHVaxx}.

\begin{figure}[ht]
\centerline{\includegraphics[height=2.5cm]{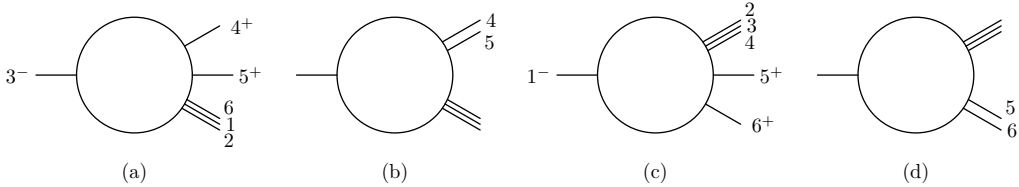} }
\caption{This set has 4 diagrams. The tensor reduction is by
using the two adjacent same helicity legs $k_{4,5}$ for $(a)+(b)$,
and $k_{5,6}$ for $(c)+(d)$.} \label{NMHVag}
\end{figure}

The 4 diagrams shown in Fig.~\ref{NMHVag} can be reduced by using the two adjacent helicity legs
$k_{4,5}$ or $k_{5,6}$ and the rational part from the first two diagrams is:
\begin{eqnarray}
R_6(1) &=&\frac 1
9((\epsilon_{1},\epsilon_{234})(\tilde{\epsilon}_6,k_1)-(\tilde{\epsilon}_6,\epsilon_{234})
(\epsilon_1,k_6) +\frac1
2(\epsilon_{234},k_6-k_1)(\tilde{\epsilon}_6,\epsilon_1))
\nonumber\\&&- \frac{\la25\ra}
{18\la45\ra}s_{345}(\epsilon_3,\epsilon_{612})
-\frac1 6\la25\ra[56](\epsilon_1,\epsilon_{234}) .
\end{eqnarray}

\begin{figure}[ht]
\centerline{\includegraphics[height=3cm]{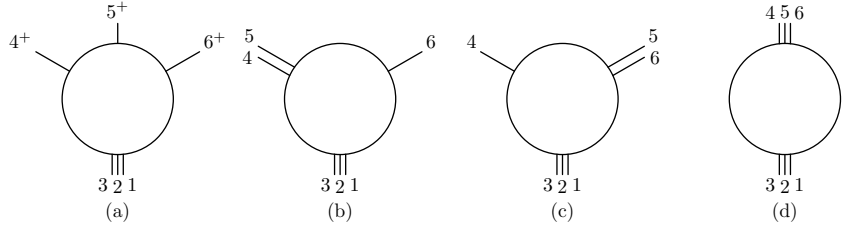} }
\caption{These diagrams have the same upper parts as the diagrams
shown in Fig.~\ref{NMHVaa}. Here the bottom part  has $k_{1,2,3}$
pinched together.} \label{NMHVai}
\end{figure}

The 4 diagrams shown in Fig.~\ref{NMHVai} have the same upper parts as the 4 diagrams shown in
Fig.~\ref{NMHVaa} apart from the bottom parts which have $k_{1,2,3}$ pinched together. The
rational part can be easily computed and we have:
\begin{eqnarray}
R_7(1) =-\frac1{6}\la25\ra[56](\epsilon_{4},\epsilon_{123})+\frac1{12}\la25\ra[6|k_{123}
\epsilon_{123}|4] .
\end{eqnarray}

\begin{figure}[ht]
\centerline{\includegraphics[height=3cm]{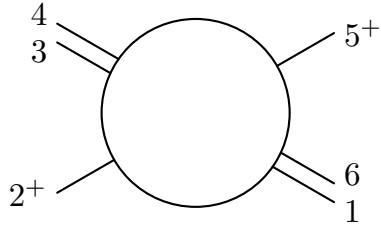} }
\caption{A two-mass-easy box diagram which is invariant under the symmetry operation
$\sigma$. } \label{NMHVaxcc}
\end{figure}

The diagram shown in Fig.~\ref{NMHVaxcc} is a two-mass-easy box diagram
which is invariant under the symmetry operation $\sigma$.
The rational part is:
\begin{eqnarray}
R_{8}(1) &=&-\frac5{18}(k_2,k_5)(\epsilon_{34},\epsilon_{61}) -
\frac1 4(\epsilon_{34},k_5)(\epsilon_{61},k_5)\, {s_{34}+s_{345}\over
s_{34}-s_{345}}
\nonumber \\ &  + & \frac4
9(\epsilon_{34},k_{5})(\epsilon_{61},k_2) -\frac1
4(\epsilon_{34},k_2)(\epsilon_{61},k_2)\, {s_{61}+s_{345}\over
s_{61}-s_{345}} .
\end{eqnarray}

\begin{figure}[ht]
\centerline{\includegraphics[height=3cm]{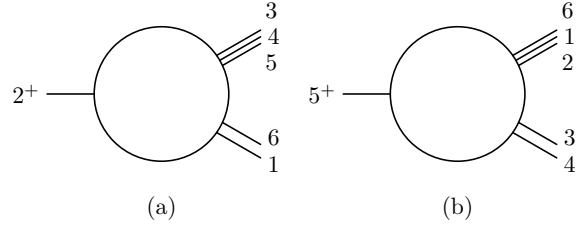} }
\caption{This set has 4 Feynman diagrams. The other 2 diagrams are
obtained by symmetry operation $\sigma$ from the above 2
diagrams.} \label{NMHVaf}
\end{figure}

The rational part from the first diagram in
Fig.~\ref{NMHVaf} is:
\begin{eqnarray}
R_{9,1}(1) &=&\frac 1 6{s_{61}+s_{345}\over
(s_{61}-s_{345})^2}(\epsilon_{2},k_{61})(\epsilon_{61},k_{2})(\epsilon_{345},k_{2})
\nonumber\\&& +\frac1{12}{s_{61}+s_{345}\over
s_{61}-s_{345}}((\epsilon_{2},\epsilon_{345})(\epsilon_{61},k_{2})
+(\epsilon_{2},\epsilon_{61})(\epsilon_{345},k_{2})) \nonumber\\&&
-\frac 1 4{(\epsilon_{2},k_{61})\over
s_{61}-s_{345}}((\epsilon_{61},k_{2})
((\epsilon_{3},\epsilon_{45})+(\epsilon_{34},\epsilon_{5}))+(\epsilon_6,\epsilon_1)
\, (\epsilon_{345},k_2)) \nonumber\\&& -\frac
1{36}\Big(9((\epsilon_{2},\epsilon_{345})(\epsilon_{6},\epsilon_{1})
+(\epsilon_{2},\epsilon_{61})(\epsilon_{3},\epsilon_{45})+(\epsilon_{2},\epsilon_{61})\,
(\epsilon_{34},\epsilon_{5})) \nonumber\\&&
+7(-(\epsilon_{2},\epsilon_{345})(\epsilon_{61},k_{2})+(\epsilon_{2},\epsilon_{61})\,
(\epsilon_{345},k_{2}))
\nonumber\\&&
+4(\epsilon_{61},\epsilon_{345})(\epsilon_{2},k_{61})\Big),
\end{eqnarray}
and the rational part from the 2nd diagram in Fig.~\ref{NMHVaf} is:
\begin{eqnarray}
R_{9,2}(1)&=&\frac 1 6{s_{34}+s_{345}\over
(s_{34}-s_{345})^2}(\epsilon_{5},k_{34})(\epsilon_{34},k_{5})(\epsilon_{612},k_{5})
\nonumber\\&& +\frac1{12}{s_{34}+s_{345}\over
s_{34}-s_{345}}((\epsilon_{5},\epsilon_{612})(\epsilon_{34},k_{5})
+(\epsilon_{5},\epsilon_{34})(\epsilon_{612},k_{5})) \nonumber\\&&
-\frac 1 4{(\epsilon_{5},k_{34})\over
s_{34}-s_{345}}((\epsilon_{34},k_{5})
((\epsilon_{6},\epsilon_{12})+(\epsilon_{61},\epsilon_{2}))+(\epsilon_3,\epsilon_4)\,
(\epsilon_{612},k_5)) \nonumber\\&& -\frac
1{36}\Big(9((\epsilon_{5},\epsilon_{612})(\epsilon_{3},\epsilon_{4})
+(\epsilon_{5},\epsilon_{34})((\epsilon_{6},\epsilon_{12})+(\epsilon_{61},\epsilon_{2})))
\nonumber\\&&
-7((\epsilon_{5},\epsilon_{612})(\epsilon_{34},k_{5})-(\epsilon_{5},\epsilon_{34})
(\epsilon_{612},k_{5})) \nonumber\\&&
+4(\epsilon_{34},\epsilon_{612})(\epsilon_{5},k_{34})\Big).
\end{eqnarray}

\begin{figure}[ht]
\centerline{\includegraphics[height=3cm]{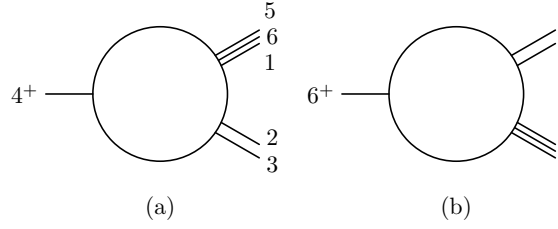} }
\caption{These 2 diagrams are related by the symmetry operation
$\sigma$.} \label{NMHVagg}
\end{figure}

The rational part from the second diagram in Fig.~\ref{NMHVagg}
is:
\begin{eqnarray}
R_{9,3}(1) & = &-\frac 1 6{s_{12}+s_{345}\over
(s_{12}-s_{345})^2}(\epsilon_{6},k_{12})(\epsilon_{12},k_{6})(\epsilon_{345},k_{6})
\nonumber\\&& -\frac1{12}{s_{12}+s_{345}\over
s_{12}-s_{345}}((\epsilon_{6},\epsilon_{345})(\epsilon_{12},k_{6})
+(\epsilon_{6},\epsilon_{12})(\epsilon_{345},k_{6})) \nonumber\\&&
-\frac 1 4{(\epsilon_{6},k_{12})(\epsilon_{12},k_{6})\over
s_{12}-s_{345}}
((\epsilon_{3},\epsilon_{45})+(\epsilon_{34},\epsilon_{5}))
\nonumber\\&& -\frac
1{36}\Big(9(\epsilon_{6},\epsilon_{12})((\epsilon_{3},\epsilon_{45})
+(\epsilon_{5},\epsilon_{34})) \nonumber\\&&
+7((\epsilon_{6},\epsilon_{345})(\epsilon_{12},k_{6})-(\epsilon_{6},\epsilon_{12})
(\epsilon_{345},k_{6})) \nonumber\\&&
-4(\epsilon_{12},\epsilon_{345})(\epsilon_{6},k_{12})\Big) .
\end{eqnarray}
Adding these three results together, we get:
\begin{equation}
R_{9}(1) = R_{9,1}(1) + R_{9,2}(1) + R_{9,3}(1).
\end{equation}

\begin{figure}[ht]
\centerline{\includegraphics[height=2.2cm]{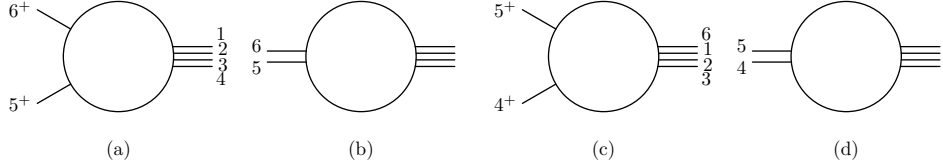} }
\caption{This set has 4 diagrams. Tensor reduction is easy with
$k_{5,6}$ or $k_{4,5}$.} \label{NMHVaj}
\end{figure}

The rational part from the first 2 Feynman diagrams of fig. \ref{NMHVaj} is:
\begin{eqnarray}
R_{10}(1) & = & \frac1 6(\epsilon_5,\epsilon_6)(\epsilon_{1234},k_5)
\nonumber \\
& - & \frac1
4(\epsilon_5,\epsilon_6)((\epsilon_1,\epsilon_{234})
+(\epsilon_{12},\epsilon_{34})+(\epsilon_{123},\epsilon_4)) .
\end{eqnarray}

There are other 8 Feynman diagrams which have the same form as given in Fig.~\ref{NMHVaj},
but with the 2 massless legs of the one-mass triangle having opposite helicities. These
give 0 by using the previous result, eq.~(\ref{eqzero}).   We will not show these diagrams
here.

\begin{figure}[ht]
\centerline{\includegraphics[height=3cm]{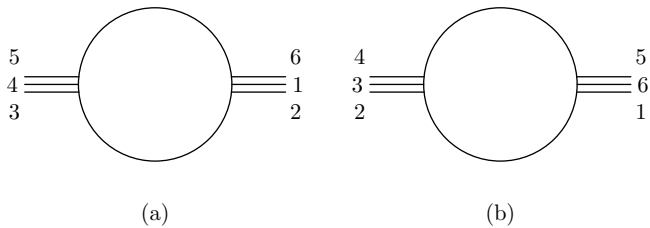} }
\caption{2 bubble diagrams.} \label{NMHVah}
\end{figure}

The last 2 diagrams are shown in Fig.~\ref{NMHVah} and the rational part is:
\begin{eqnarray}
R_{11}(1) = -\frac1 9 s_{345}(\epsilon_{345},\epsilon_{612}) .
\end{eqnarray}

Combining all the above results together, the complete rational
part is given as follows:
\begin{equation}
R = F\, \sum_{g\in G} R(g) = F\, \sum_{g\in G} \sum_{i=1}^{11} R_i(g),
\end{equation}
where
\begin{equation}
F = -{1\over  [1\,2]\, [2\,3] \, \langle2\,5\rangle^2
\, \langle4\,5\rangle\,\langle5\,6\rangle },
\end{equation}
is an overall factor from the omitted factors of the polarization
vectors.

The above result was also cross-checked with the explicit results
of Berger, Bern, Dixon, Forder and Kosower \cite{BDKE} and we
found exact (rational number) agreement.

\section{NMHV: $R(1^-2^-3^+4^-5^+6^+)$}

Starting from this section we will compute the rational parts for the two non-MHV helicity
configurations. These will give much more complicated algebraic expressions if one writes down the
complete expressions. The strategy we use is to use another level of definition
and express the result either explicitly or implicity through the (rational parts of the)
2-mass and 3-mass triangle integrals, and the 2-mass-hard box integrals. The explicit results
in terms of the more elementary quantities, i.e., spinor products and kinematic variables,
are collected in an appendix for quick reference.

For the NMHV($1^-2^-3^+4^-5^+6^+$), the symmetry group is $G=\{1,
\sigma\}$ where $\sigma = \{ i  \leftrightarrow 7-i\}$ followed by
conjugation. We can choose the following polarization vectors:
\begin{eqnarray}
\epsilon_1   =  {\lambda_1\tilde\lambda_2 \over [1\,2]}, & &
\epsilon_6   =  {\lambda_5\tilde\lambda_6 \over \langle 5\,6 \rangle }, \\
\epsilon_2   =  {\lambda_2\tilde\lambda_1 \over [2\,1]}, & &
\epsilon_5   =  {\lambda_6\tilde\lambda_5 \over \langle 6\,5 \rangle}, \\
\epsilon_4   =  {\lambda_4\tilde\lambda_2 \over [4\,2]}, & &
\epsilon_3   =  {\lambda_5\tilde\lambda_3 \over \langle 5\,3
\rangle} .
\end{eqnarray}
We note again that we will omit the overall factors for all
polarization vectors.

\begin{figure}[ht]
\centerline{\includegraphics[height=2cm]{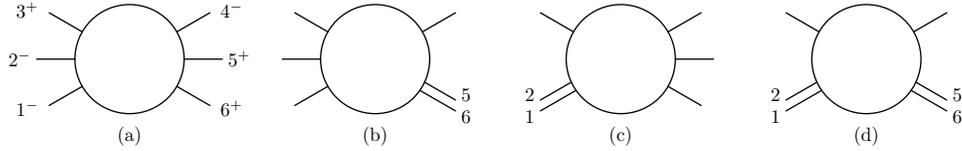} } \caption{Diagrams (a) and (d) are
invariant under the symmetry operation $\sigma$. Diagram (b) changes to diagram (c) under
the symmetry operation $\sigma$. }
\label{NMHVTwoa}
\end{figure}

The first set of diagrams are given in Fig.~\ref{NMHVTwoa} which
includes the only hexagon Feynman diagram. The tensor reduction
can be done either with $k_{5,6}$ or $k_{1,2}$, but we can not do
tensor reduction with both for all terms. By doing tensor
reduction firstly with $k_{5,6}$, we have:
\begin{eqnarray}
R_1 & = &  \left( - I^{(1)}\, I^{(5)} + \frac{1}{2} \, I^{(5)}\,
I^{(6)}  +
 \frac{1}{2} \, I^{(1)}\, I^{(6)} \right) \,
  (\epsilon_3,p-k_{12})(\epsilon_4,p-k_{123}) \nonumber \\
 &  \times &  \left(  (\epsilon_1,p)
 (\epsilon_2,p-k_1) +  \, I^{(2)} \, (  (\epsilon_{12},p) + {s_{12}\over2})
 \right) ,
\end{eqnarray}
where $I^{(i)}$ denotes the inverse propagator between $k_{i-1}$ and $k_i$.
The last factor can be reduced further with the external momentum
$k_2$ by noting the following relation:
\begin{equation}
 (\epsilon_1,p)
 (\epsilon_2,p-k_1) +  \, I^{(2)} \, (  (\epsilon_{12},p) + {s_{12}\over2})
 = \frac{I^{(2)}}{2}(k_{12}, p - {k_{12}\over2}) - I^{(3)}(k_1,p) .
\end{equation}
By using these results the rational part can be computed easily. We have:
\begin{eqnarray}
R_1 & = & I_3^{3m}(k_1,\epsilon_3,\epsilon_4) + (\epsilon_4,k_5)\,
 I_3^{3m}(k_1,\epsilon_3 ) \nonumber \\
& - & {1\over 2} \, \left[ \tilde
I_3^{2m(4)}(\epsilon_4,k_1,\epsilon_3) + (s_{56}-s_{234})\,
\tilde I_3^{2m(4)}(\epsilon_4 ,\epsilon_3) \right] \nonumber \\
& - & {1\over 2} \, \left[
I_3^{2m(3)}(\epsilon_3,\epsilon_4,k_{12}) + {s_{12} } \,
I_3^{2m(3)}(\epsilon_3 ,\epsilon_4) \right] \nonumber \\
& + & {1\over 36} \, s_{34}\,(\epsilon_{34},k_{12}) - {1\over 36}
( (\epsilon_3,k_2)(\epsilon_4,k_3) - 2
s_{23}(\epsilon_3,\epsilon_4))
\nonumber \\
& - &  {1\over 72}( (\epsilon_3,k_{12})(\epsilon_4,k_{56})+ 2
s_{123}(\epsilon_3,\epsilon_4)) .
\end{eqnarray}

\begin{figure}[ht]
\centerline{\includegraphics[height=3cm]{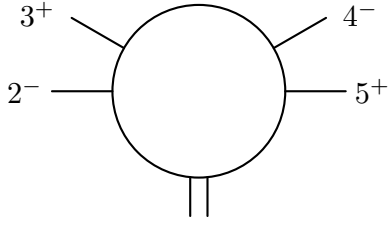} } \caption{This
diagram is invariant under the symmetry operation $\sigma$:
$1\leftrightarrow6,~2\leftrightarrow5,~3\leftrightarrow4$ (with
conjugation).} \label{NMHVTwob}
\end{figure}

The diagram shown in Fig.~\ref{NMHVTwob} is invariant under the symmetry operation $\sigma$.
Its tensor reduction can be done for $k_2$ and $k_5$ because of our special choice of the
reference momenta for $\epsilon_{3,4}$. We have:
\begin{eqnarray}
R_2 & = & - \langle4\,3\rangle [3\,2]\, (I^{(5)} - I^{(6)})\,
I^{(1)}\,
(\lambda_2\tilde\lambda_1,p) \, (\lambda_6\tilde\lambda_3,p)(\epsilon_{61},p) \nonumber \\
& + &  \langle5\,4\rangle [4\,3]\, (I^{(2)} - I^{(3)} )\,
I^{(1)}\,
(\lambda_4\tilde\lambda_1,p) \, (\lambda_6\tilde\lambda_5,p)(\epsilon_{61},p) \nonumber \\
& + & ( I^{(2)}\, (\lambda_4\tilde\lambda_1,p-k_{12}) - I^{(3)} \,
(\lambda_4\tilde\lambda_1,p-k_{1}) ) \nonumber \\
& & \times ( I^{(5)}\, (\lambda_6\tilde\lambda_3,p+k_{6})   -
I^{(6)}  \, (\lambda_6\tilde\lambda_3,p+k_{56}) ) \, I^{(1)} \,
P_{61}(p) ,
\end{eqnarray}
where
\begin{equation}
P_{61}(p) =    (\epsilon_{61} , p+k_6) -
\frac{1}{2}(\epsilon_6,\epsilon_1) .
\end{equation}
The rational part is computed by using the
formulas for box and triangle integrals and we have:
\begin{eqnarray}
R_2 & = & - \langle4\,3\rangle [3\,2] (
I_4^{2mh(2)}(\lambda_2\tilde\lambda_1, \lambda_6\tilde\lambda_3,
\epsilon_{61}) - I_4^{1m(2)}(\lambda_2\tilde\lambda_1,
\lambda_6\tilde\lambda_3, \epsilon_{61}) ) \nonumber \\
& + &  \langle5\,4\rangle [4\,3]\, ( I_4^{1m(3)}(
\lambda_4\tilde\lambda_1, \lambda_6\tilde\lambda_5, \epsilon_{61})
- I_4^{2mh(4)}(\lambda_4\tilde\lambda_1, \lambda_6\tilde\lambda_5,
\epsilon_{61}) )
\nonumber \\
& + & I_3^{2m(3)}(\lambda_6\tilde\lambda_3,\epsilon_{61},
\lambda_4\tilde\lambda_1) -  \langle6|k_{45}|3] \,
I_3^{2m(3)}(\lambda_4\tilde\lambda_1, \epsilon_{61})
\nonumber \\
& - & ((\epsilon_{61},k_{45}) + {1\over2}(\epsilon_6,\epsilon_1))
\,I_3^{2m(3)}(\lambda_6\tilde\lambda_3,\lambda_4\tilde\lambda_1)
\nonumber \\
& + &  \tilde I_3^{2m(4)}(\lambda_4\tilde\lambda_1,
\lambda_6\tilde\lambda_3,\epsilon_{61}) + \langle4|k_{23}|1] \,
\tilde I_3^{2m(4)}(\lambda_6\tilde\lambda_3, \epsilon_{61})
\nonumber \\
& + & ((\epsilon_{61},k_{23}) - {1\over2}(\epsilon_6,\epsilon_1))
\,\tilde
I_3^{2m(4)}(\lambda_6\tilde\lambda_3,\lambda_4\tilde\lambda_1)
\nonumber \\
& - &
I_3^{3m}(\lambda_4\tilde\lambda_1,\epsilon_{61},\lambda_6\tilde\lambda_3)
- ((\epsilon_{61},k_{23}) - {1\over2}(\epsilon_6,\epsilon_1))
\, I_3^{3m}(\lambda_4\tilde\lambda_1,\lambda_6\tilde\lambda_3) \nonumber \\
& + & {7\over 18}\, s_{34}(\tilde\epsilon_{34},\epsilon_{61}) +
{1\over 4}\, \langle6\,4\rangle\, [1\,3] \,
(\epsilon_{61},k_3-k_4) ,
\end{eqnarray}
where
\begin{eqnarray}
\tilde\epsilon_{34} & = & \epsilon_{34}|_{\epsilon_3 \to
\lambda_6\tilde\lambda_3, \epsilon_4 \to
\lambda_4\tilde\lambda_1} \nonumber \\
& = & {\langle6\,4\rangle\over\langle3\,4\rangle}\,
\lambda_4\tilde\lambda_1 - {[1\,3]\over [4\,3]}\,
\lambda_6\tilde\lambda_3 + {\langle6\,4\rangle \, [1\,3] \over 2
\,  \langle 3\,4\rangle\, [4\,3]} \, (k_3-k_4) .
\end{eqnarray}

\begin{figure}[ht]
\centerline{\includegraphics[height=2.3cm]{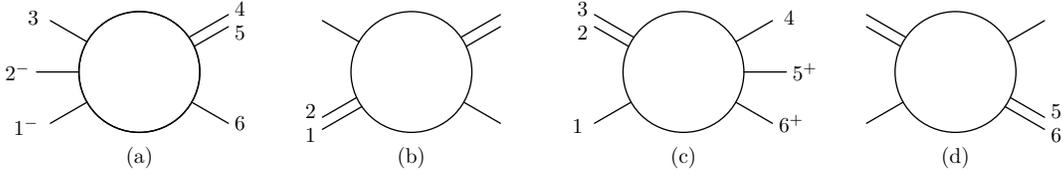} } \caption{4
Feynman diagrams which can be reduced easily with $k_{1,2}$ ((a)
and (b)) or $k_{5,6}$ ((c) and (d)).  (c) and (d) are obtained
from (a) and (b) by symmetry operation $\sigma$.} \label{NMHVTwoc}
\end{figure}

The four Feynman diagrams in Fig.~\ref{NMHVTwoc} are related by the
symmetry operation $\sigma$. The rational part from diagrams (a)
and (b) is:
\begin{eqnarray}
R_3(1) & = &  {1\over2}\,
I_3^{2m(3)}(\epsilon_3,\epsilon_{45},\epsilon_6) + {1\over4}\,
I_3^{2m(3)}(\epsilon_3,2 (\epsilon_6,k_{12}) \epsilon_{45} -
(\epsilon_4,\epsilon_5)
\, \epsilon_6) \nonumber \\
& + & {1\over2}\, \tilde I_3^{2m(6)}(\epsilon_6,
\epsilon_3,\epsilon_{45}) - {1\over4}\, \tilde
I_3^{2m(6)}(\epsilon_6,2 (\epsilon_3,k_{12}) \epsilon_{45} +
(\epsilon_4,\epsilon_5) \, \epsilon_3) \nonumber \\
& - &  I_3^{3m}(\epsilon_{45},\epsilon_3,\epsilon_6) + \left(
(\epsilon_{45},k_{23})+ {1\over 2}\, (\epsilon_4,\epsilon_5)
\right) \, I_3^{3m}( \epsilon_3,\epsilon_6) .
\end{eqnarray}
The rational part from diagrams (c) and (d) is denoted as
$R_3(\sigma)$. It is obtained from $R_3(1)$ by the symmetry
operation $\sigma$.

\begin{figure}[ht]
\centerline{\includegraphics[height=2.5cm]{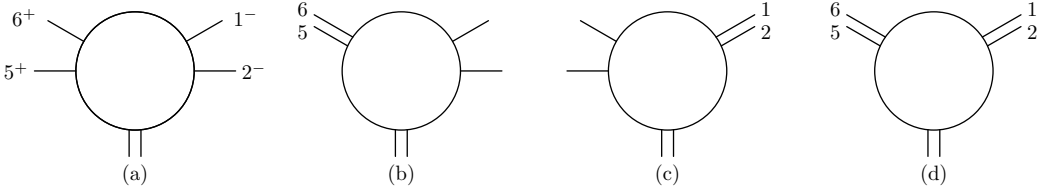} }
\caption{These 4 diagrams have the same structure as those in
Fig.~\ref{NMHVTwoa} with $k_{3,4}$ replaced by $k_{34}$.  }
\label{NMHVTwod}
\end{figure}

For the 4 diagrams shown in Fig.~\ref{NMHVTwod},
the rational part can be obtained by using the same tensor
reduction techniques as for Fig.~\ref{NMHVTwoa} and the rational
part is:
\begin{eqnarray}
R_4&=& \frac 1{36}
((\epsilon_{34},k_2)(k_1,k_{56})-(\epsilon_{34},k_5)(k_6,k_{12}))
\nonumber \\& + & \frac 1{18}((\epsilon_{34},k_1)(k_5,k_{61})
-(\epsilon_{34},k_6)(k_2,k_{61})) \nonumber \\&  - & \frac {1}{72}
\, s_{34} \, ( \epsilon_{34},k_{12} -  k_{56})
-\frac{1}{6}(s_{12}(\epsilon_{34},\epsilon_{56}) +
s_{56}(\epsilon_{34},\epsilon_{12})) .
\end{eqnarray}

\begin{figure}[ht]
\centerline{\includegraphics[height=2cm]{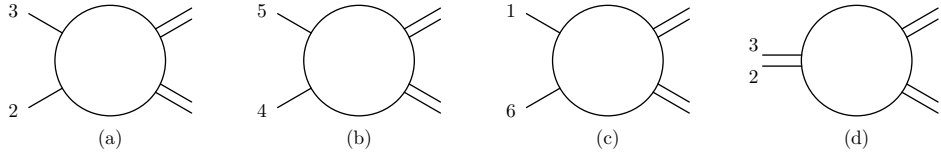} } \caption{Four
diagrams which should be computed individually.  }
\label{NMHVTwoe}
\end{figure}

The rational part for the 3 two-mass-hard box diagrams and the three-mass
 triangle diagram is:
\begin{eqnarray}
R_{5} & = &
I_4^{2mh(2)}(\lambda_2\tilde\lambda_1,\lambda_5\tilde\lambda_3,\epsilon_{45},
\epsilon_{61}; -(\epsilon_4,\epsilon_5)/2,
-(\epsilon_6,\epsilon_1)/2)
\nonumber \\
& + &
I_4^{2mh(4)}(\lambda_4\tilde\lambda_2,\lambda_6\tilde\lambda_5,\epsilon_{61},
\epsilon_{23}; -(\epsilon_6,\epsilon_1)/2,
-(\epsilon_2,\epsilon_3)/2)
\nonumber \\
& + &
I_4^{2mh(6)}(\lambda_5\tilde\lambda_6,\lambda_1\tilde\lambda_2,\epsilon_{23},
\epsilon_{45}; -(\epsilon_2,\epsilon_3)/2,
-(\epsilon_4,\epsilon_5)/2)
\nonumber \\
& + & I_3^{3m}(\epsilon_{23},\epsilon_{45},\epsilon_{61}) -
{1\over2}(\epsilon_2,\epsilon_3)
\, I_3^{3m}(\epsilon_{45},\epsilon_{61}) \nonumber \\
& - &   {1\over2}(\epsilon_4,\epsilon_5) \,
I_3^{3m}(\epsilon_{61},\epsilon_{23}) -
{1\over2}(\epsilon_6,\epsilon_1) \,
I_3^{3m}(\epsilon_{23},\epsilon_{45}) ,
\end{eqnarray}
just by using the definition of $I_4^{2mh(i)}$ and $I^{3m}_3$.
See the appendix for explicit formulas.

\begin{figure}[ht]
\centerline{\includegraphics[height=3cm]{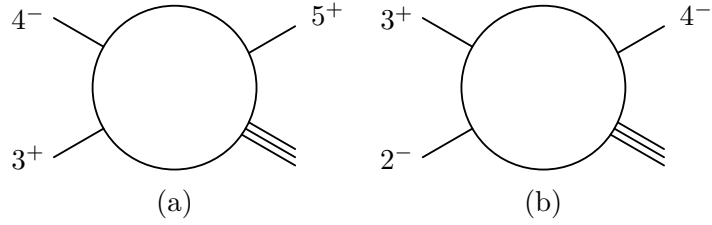} } \caption{2
Feynman diagrams. The 2nd can be obtained from the first one by
the symmetry operation $\sigma$. } \label{NMHVTwof}
\end{figure}

The first diagram in Fig.~\ref{NMHVTwof} gives the following
rational part:
\begin{eqnarray}
R_6(1) & = & {\langle 6\,5\rangle \over \langle 3\,5\rangle }\,
\left[ {4\over 9}\, ( (\epsilon_4,k_3)(\epsilon_{612},k_5) +
(\epsilon_4,k_5)(\epsilon_{612},k_3) )   \right.\nonumber \\
& & - {5\over 9}\, (k_3,k_5)(\epsilon_4,\epsilon_{612}) +
{1\over4}\,( (\epsilon_4,k_3)(\epsilon_{612},k_3) +
(\epsilon_4,k_5)(\epsilon_{612},k_5) )
 \nonumber \\
& & +  \left. {1\over4} \left[ {s_{45} + s_{345}\over s_{45} -
s_{345}} \, (\epsilon_4,k_3)(\epsilon_{612},k_3) + {s_{34} +
s_{345}\over s_{34} - s_{345}} \,
(\epsilon_4,k_5)(\epsilon_{612},k_5) \right] \right ]
\nonumber \\
& + & {\langle 6\,3\rangle \over \langle 5\,3\rangle }\, \left[ -
{1\over9}\, s_{34}(\epsilon_{34}, \epsilon_{612}) +
I_3^{2m(3)}(\epsilon_3,\epsilon_4,\epsilon_{612}) \right.
\nonumber \\
& & \left.  - {1\over2}\,
((\epsilon_6,\epsilon_{12})+(\epsilon_{61},\epsilon_{2}))\,
I_3^{2m(3)}(\epsilon_3,\epsilon_4) \right]  .
\end{eqnarray}
The rational part of the other diagram Fig.~\ref{NMHVTwof} is
$R_6(\sigma)$ and it is obtained from  $R_6(1)$ by the symmetry
operation $\sigma$.

\begin{figure}[ht]
\centerline{\includegraphics[height=2cm]{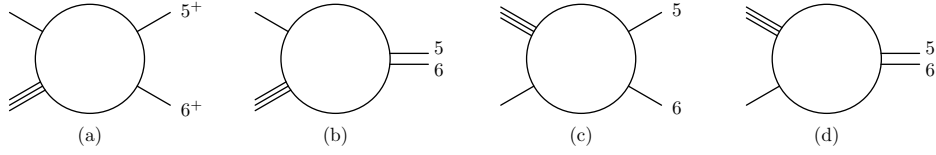} } \caption{This
set has 8 Feynman diagrams. Four more diagrams are obtained from
these four diagrams by the symmetry operation $\sigma$. The above
four diagrams  can be reduced easily with $k_{5,6}$.}
\label{NMHVTwog}
\end{figure}

The four diagrams shown in Fig.~\ref{NMHVTwog} give the following
rational part:
\begin{eqnarray} R_7(1) & = & {1\over 18}\,
(\epsilon_4,k_5)\,(\epsilon_{123},k_6) + {1\over 18}\, (
2 s_{45} - s_{123} - 3 s_{56} ) \, (\epsilon_{123},\epsilon_4) \nonumber \\
& + &    {1\over 18}\, (\epsilon_1,k_6)\,(\epsilon_{234},k_5) +
{1\over 18}\, ( 2 s_{61} - s_{234} - 3 s_{56} ) \,
(\epsilon_{1},\epsilon_{234}) .
\end{eqnarray}
$R_7(\sigma)$ is obtained from $R_7(1)$ by the symmetry operation
$\sigma$ (with conjugation).

\begin{figure}[ht]
\centerline{\includegraphics[height=2cm]{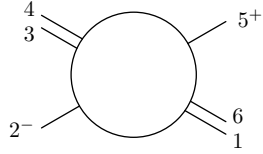} } \caption{A
two-mass-easy box diagram. } \label{NMHVTwom}
\end{figure}

The rational part from the diagram shown in Fig.~\ref{NMHVTwom}
decomposes as $R_8(1) + R_8(\sigma)$ and we have:
\begin{eqnarray}
R_{8}(1) & = & {[1\,2]\langle6\,5\rangle \over
[5\,2]\langle2\,5\rangle} \left[  {5 \over 18}\,
\langle2|\epsilon_{34}|5]\,  \langle2|\epsilon_{61}|5]
 + {\langle2|k_{34}|5]^2 \, (\epsilon_{34},k_2)\,(\epsilon_{61},
 k_2) \over
3\,  \, (s_{34}-s_{234})\, ( s_{61}-s_{345}) }
 \right]
\nonumber \\
& + & {\langle6\,2\rangle \over\langle5\,2\rangle}\, \left[ \tilde
I_3^{2m(2)}(\lambda_2\tilde \lambda_1,
\epsilon_{34},\epsilon_{61})- {1\over2} \tilde
I_3^{2m(2)}(\lambda_2\tilde \lambda_1,
(\epsilon_6,\epsilon_1)\epsilon_{34} +
(\epsilon_3,\epsilon_4)\epsilon_{61})
\right. \nonumber \\
& & - \left.  I_3^{2m(2)}(\lambda_2\tilde \lambda_1,
\epsilon_{34},\epsilon_{61}) +  {1\over2}\,
I_3^{2m(2)}(\lambda_2\tilde \lambda_1,
(\epsilon_6,\epsilon_1)\epsilon_{34} +
(\epsilon_3,\epsilon_4)\epsilon_{61})
\right] \nonumber  \\
& -  & {[1\,5]\langle6\,2\rangle \over 18 \,
[2\,5]\langle5\,2\rangle} \, ( (k_2,k_5)(\epsilon_{34},
\epsilon_{61}) - (\epsilon_{34},k_2)(\epsilon_{61},k_5)) ,
\end{eqnarray}
whereas $R_8(\sigma)$ is obtained from $R_8(1)$ by the symmetry
operation $\sigma$ (with conjugation).

\begin{figure}[ht]
\centerline{\includegraphics[height=2cm]{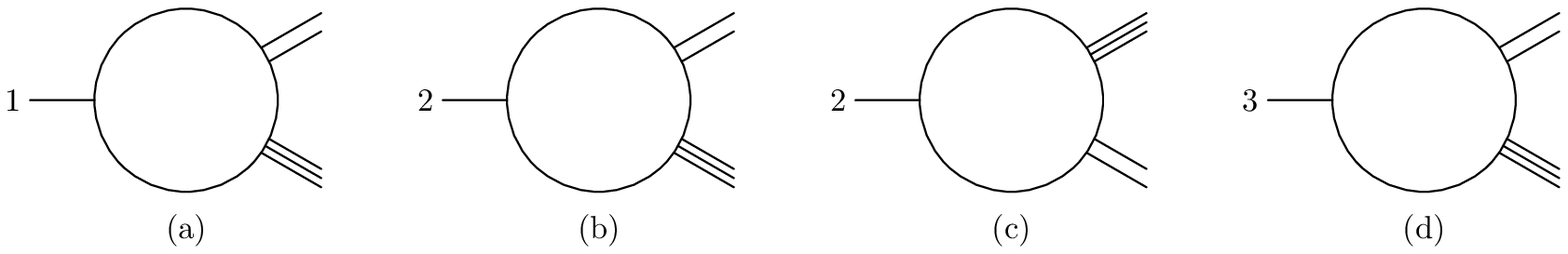} } \caption{Four
2-mass triangle diagrams which should be computed individually.  }
\label{NMHVTwoh}
\end{figure}

The diagrams in Fig.~\ref{NMHVTwoh} give the following rational
part:\begin{eqnarray} R_9(1) & = &
I_3^{2m(1)}(\epsilon_1,\epsilon_{23}, \epsilon_{456}) +
I_3^{2m(3)}(\epsilon_3,\epsilon_{45}, \epsilon_{612}) \nonumber \\
& + &
I_3^{2m(2)}(\epsilon_2,\epsilon_{34}, \epsilon_{561}) +
\tilde I_3^{2m(2)}(\epsilon_2,\epsilon_{345}, \epsilon_{61}) \nonumber
\\
& - &
{1\over2}\, I_3^{2m(1)}(\epsilon_1,
(\epsilon_2,\epsilon_3)\epsilon_{456}+
( (\epsilon_4,\epsilon_{56}) + (\epsilon_{45},\epsilon_6) ) \epsilon_{23}) \nonumber \\
& - & {1\over2}\, I_3^{2m(3)}(\epsilon_3,
(\epsilon_4,\epsilon_5)\epsilon_{612}+
( (\epsilon_6,\epsilon_{12}) + (\epsilon_{61},\epsilon_2) ) \epsilon_{45}) \nonumber \\
& - & {1\over2}\, I_3^{2m(2)}(\epsilon_2,
(\epsilon_3,\epsilon_4)\epsilon_{561}+
( (\epsilon_5,\epsilon_{61}) + (\epsilon_{56},\epsilon_1) ) \epsilon_{34}) \nonumber \\
& - & {1\over2}\, \tilde I_3^{2m(2)}(\epsilon_2,
((\epsilon_3,\epsilon_{45}) + (\epsilon_{34},\epsilon_{5}))
\epsilon_{61} +
  (\epsilon_6,\epsilon_{1})   \epsilon_{345})  .
\end{eqnarray}

\begin{figure}[ht]
\centerline{\includegraphics[height=2cm]{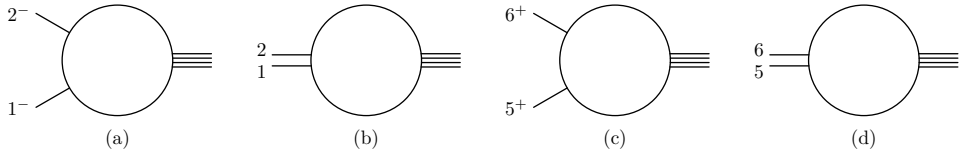} } \caption{Four
1-mass triangle diagrams which can be computed either directly or
by tensor reduction with $k_{1,2}$ or $k_{5,6}$.} \label{NMHVTwoj}
\end{figure}

The rational part from the first two diagrams in Fig.~\ref{NMHVTwoj} is:
\begin{equation}
R_{10}(1)  =
 {1\over 4}\, s_{12}\, ( (\epsilon_{3}, \epsilon_{456})
 + (\epsilon_{34},\epsilon_{56}) +
 (\epsilon_{3},\epsilon_{456}) ) -
   {1 \over 6} \, s_{12} (\epsilon_{3456}, k_1)  .
\end{equation}

\begin{figure}[ht]
\centerline{\includegraphics[height=3cm]{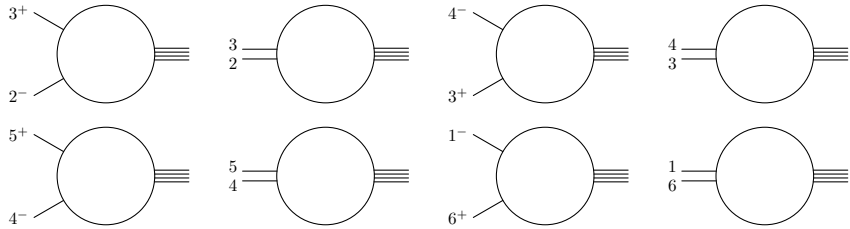} }
\caption{These 8 diagrams give a vanishing contribution to the
rational part.} \label{NMHVTwol}
\end{figure}

\begin{figure}[ht]
\centerline{\includegraphics[height=2.5cm]{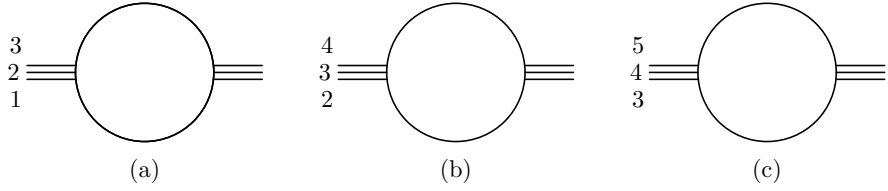} }
\caption{The three remaining bubble diagrams.} \label{NMHVTwon}
\end{figure}

The last contribution to the rational part is from the remaining
three bubble diagrams shown in Fig.~\ref{NMHVTwon} and it is given
by:
\begin{equation}
R_{11} = - {1\over 9}( s_{123}(\epsilon_{123},\epsilon_{456})
 +s_{234}(\epsilon_{234},\epsilon_{561}) + s_{345}(\epsilon_{345},\epsilon_{612}) ) .
\end{equation}

The complete result for the rational part is:
\begin{eqnarray}
R  = F \, \left[ \sum_{i=1,2,4,5,11} R_i + \sum_{i=3, 6,7,8,
9,10}\sum_{g } R_i(g) \right] ,
\end{eqnarray}
where
\begin{equation}
F =   { 1\over \langle3\,5\rangle\,[2\,4] \, ([1\,2] \,
\langle5\,6\rangle)^2},
\end{equation}
is an overall factor.

\section{NMHV: $R(1^-2^+3^-4^+5^-6^+)$}

In this case we choose the following polarization vectors:
\begin{eqnarray}
\epsilon_1   =  {\lambda_1\tilde\lambda_5 \over [1\,5]}, & &
\epsilon_2   =  {\lambda_6\tilde\lambda_2 \over \langle 6\,2 \rangle }, \\
\epsilon_3   =  {\lambda_3\tilde\lambda_1 \over [3\,1]}, & &
\epsilon_4   =  {\lambda_2\tilde\lambda_4 \over \langle 2\,4 \rangle}, \\
\epsilon_5   =  {\lambda_5\tilde\lambda_3 \over [5\,3]}, & &
\epsilon_6   =  {\lambda_4\tilde\lambda_6 \over \langle 4\,6
\rangle} .
\end{eqnarray}
This helicity configuration has the biggest symmetry group which
is $Z_6$ generated by $\tau: i \to i+1$ accompanied by
conjugation. Because of this symmetry group, we need only to
compute a few diagrams from each set and the complete result will
be obtained by a summation over the symmetry group. For an
invariant diagram, tensor reduction often (partially) destroys
the symmetry and the symmetry is not manifest in the resulting
expression of the rational part. In this case it is often simpler
to give the complete result for this invariant diagram as a whole
or as a summation over only a subgroup of the symmetry group
$Z_6$.

Let us first  start with the  diagram (a) in
Fig.~\ref{FeynmanSixPoint}. We can do the tensor reduction with
$k_{1,4}$ and we have
\begin{eqnarray}
R_1 & = & -\langle4\,5\rangle[5\,6] (I^{(1)}-I^{(2)})\,
(I^{(3)}-I^{(4)})\,(I^{(5)}-I^{(6)})\,
(\lambda_2\tilde\lambda_4,p)(\lambda_6\tilde\lambda_2,p) \nonumber \\
& - & \langle3\,2\rangle[2\,1] (I^{(2)}-I^{(3)})\,
(I^{(4)}-I^{(5)})\,(I^{(6)}-I^{(1)})\,
(\lambda_5\tilde\lambda_3,p)(\lambda_1\tilde\lambda_5,p) \nonumber \\
& - & I^{(1)}\, I^{(5)} \,
(\lambda_3\tilde\lambda_5,p-k_1)(\lambda_6\tilde\lambda_2,p-k_{12})
\,(\lambda_2\tilde\lambda_6,p-k_{123})\,(\lambda_5\tilde\lambda_3,p+k_6) \nonumber \\
& - & I^{(2)}\, I^{(4)} \,
(\lambda_2\tilde\lambda_6,p+k_{56})(\lambda_5\tilde\lambda_3,p+k_6)
\,(\lambda_3\tilde\lambda_5,p )\,(\lambda_6\tilde\lambda_2,p-k_{12})\nonumber \\
& + & I^{(1)}\, I^{(4)} \,
(\lambda_3\tilde\lambda_5,p-k_1)(\lambda_6\tilde\lambda_2,p-k_{12})
\,(\lambda_2\tilde\lambda_6,p+k_{56})\,(\lambda_5\tilde\lambda_3,p+k_6) \nonumber \\
& + & I^{(2)}\, I^{(5)} \, (\lambda_6\tilde\lambda_2,p-k_{12})
\,(\lambda_2\tilde\lambda_6,p-k_{123})\,(\lambda_5\tilde\lambda_3,p+k_6)
\,(\lambda_3\tilde\lambda_5,p )  . \label{firstzsix}
\end{eqnarray}
We denote the contribution (to the rational part) of each term by
$R_1^{(i)}$, $i=1,\cdots,6$. Let us compute the various terms
explicitly. We have
\begin{eqnarray}
R_{1}^{(1)} & = & -\langle4\,5\rangle\,[5\,6]\left[
 {1\over 2}\left[ {\langle2|k_{56}|4] \over
s_{56}-s_{456}} -  {\langle2|k_{23}|4] \over s_{23}-s_{234}}
\right] \, (\lambda_6\tilde\lambda_2,k_4)
 \right. \nonumber \\
& &  + I_3^{3m}(\lambda_2\tilde\lambda_4,\lambda_6\tilde\lambda_2)
-  \tilde
I_3^{3m}(\lambda_2\tilde\lambda_4,\lambda_6\tilde\lambda_2)
 \nonumber \\
& & \left.
 +  {1\over2}\left[  {\langle6|k_{12}|2] \over s_{12}-s_{345}} -
 {\langle6|k_{45}|2] \over s_{45}-s_{123}} \right] \, (\lambda_2\tilde\lambda_4,k_6) \right].
\end{eqnarray}
$R^{(2)}_1$ can be obtained  from $R_1^{(1)}$ by the symmetry
operation $\tau^3$, i.e.:
\begin{equation}
 R_1^{(2)} =
(R_1^{(1)}|_{i \to i+3})^*,
\end{equation}
where $*$ denotes conjugation: $\langle \, \rangle \leftrightarrow
[\,]$.

The contribution from the 3rd term in eq.~ (\ref{firstzsix}) is:
\begin{eqnarray}
R_1^{(3)} & = & - {1\over 6}\, (k_2,k_6)(k_3,k_5) -
\langle3|k_2|5]\, I_3^{2m(3)}(\lambda_5\tilde\lambda_3,k_6)
\nonumber \\
& + & \langle5|k_4|3]\, I_3^{2m(3)}(\lambda_3\tilde\lambda_5,k_6)
- I_3^{2m(3)} (\lambda_3
\tilde\lambda_5,\lambda_5\tilde\lambda_3,k_6) \nonumber \\
& + & (k_6,k_{23})\,
I_3^{3m}(\lambda_3\tilde\lambda_5,\lambda_5\tilde\lambda_3) +
 I_3^{3m}(\lambda_3\tilde\lambda_5,k_6,\lambda_5\tilde\lambda_3) \nonumber \\
& + &  \langle2|k_3|6]\, \left[ I_3^{3m}(\lambda_6
\tilde\lambda_2,k_5) - {1\over2}\, \left[ \langle6|k_5|2] + {
\langle6|k_1|2] \, (k_2,k_5)\over s_{61}-s_{345}}\right] \right] .
\end{eqnarray}
$R^{(4)}_1$ can be obtained  from $R_1^{(3)}$ by  the symmetry
operation $\tau^3$ just as for the term $R^{(2)}_1$ from
$R^{(1)}_1$.

The rational parts from the last two terms in
eq.~(\ref{firstzsix}) are
\begin{eqnarray}
R_1^{(5)} & = & {1\over2}(k_3,k_5)\,(k_2,k_6)+{1\over18}\,( 2
(k_3,k_6)(k_2,k_5) \nonumber \\
& & - (k_5,k_3)
(k_2,k_6)- (k_5,k_6)(k_2,k_3) ) \nonumber \\
& - & {1\over2}\, \left[ {\langle2|k_{34}k_6k_{61}|2] \over
s_{61}-s_{345} } -
{\langle2|k_{34}k_6 k_{34} |2] \over s_{34}-s_{234} } \right] \, (k_3,k_2) \nonumber \\
& - & {1\over2}\, \left[ {\langle3|k_{61}k_5k_{34}|3] \over
s_{34}-s_{345} } - {\langle3|k_{61}k_5 k_{61} |3] \over
s_{61}-s_{234} } \right] \, (k_6,k_5) , \\
 R_1^{(6)}  & = &
{1\over2}(k_3,k_5)\,(k_2,k_6)+{1\over18}\,( 2
(k_3,k_6)(k_2,k_5)  \nonumber \\
& &  - (k_5,k_3)
(k_2,k_6)- (k_5,k_6)(k_2,k_3) ) \nonumber \\
& + & {1\over2}\, \left[ {\langle2|k_{45}k_6(k_{45}-k_{12})|2]
\over s_{45}-s_{123} } -
{\langle2|(k_{45}-k_{12})k_6 k_{12} |2] \over s_{12}-s_{345} }
\right] \, (k_5,k_6) \nonumber \\
& + & {1\over2}\, \left[ {\langle5|k_{45}k_3(k_{45}-k_{12})|5]
\over s_{45}-s_{345} } - {\langle5|(k_{45}-k_{12})k_3 k_{12} |5]
\over s_{12}-s_{123} } \right] \, (k_2,k_3)  .
\end{eqnarray}

In summary the rational part from the six-point diagram (a) in
Fig.~\ref{FeynmanSixPoint} is:
\begin{equation}
R_1 = \sum_{i=1}^6 R_1^{(i)} .
\end{equation}
It is manifestly invariant only under a subgroup of the symmetry
group, because the reduction procedure breaks the symmetry group
and keeps only a subgroup manifest. Of course, $R_1$ must have a
hidden symmetry as the final result does not depend on how one
proceeds with the tensor reduction. We have checked that $R_1$ do
have the full $Z_6$ symmetry.

For the diagram (b) with $i=2$ in Fig.~\ref{FeynmanSixPoint}, the
computation proceeds much as the computation of the 6-point
diagram as given in the above. The result is:
\begin{eqnarray}
R_2(1) & = & -
I_4^{2mh(4)}(\lambda_6\tilde\lambda_4,\lambda_5\tilde\lambda_3,\epsilon_{61},
\lambda_3\tilde\lambda_1; -(\epsilon_6,\epsilon_1)/2,-\langle3|k_2|1]) \nonumber \\
& - & \langle6|k_{23}|4]\, I_4^{2mh(4)}(
\lambda_5\tilde\lambda_3,\epsilon_{61},
\lambda_3\tilde\lambda_1) \nonumber \\
& + & {[1\,3]\over [5\,3] }\, \left[ {4\over9}\, (
\langle6|k_3|4]\, (\epsilon_{61},k_5) +  \langle6|k_5|4]\,
(\epsilon_{61},k_3)
 ) \right. \nonumber \\
& & - {5\over9}\,(k_3,k_5)\, \langle6|\epsilon_{61}|4] +
{1\over4}\, ( \langle6|k_3|4]\, (\epsilon_{61},k_3) +
\langle6|k_5|4]\, (\epsilon_{61},k_5) ) \nonumber \\
& & \left. + {1\over4}\, \left[ {s_{45}+s_{345}\over
s_{45}-s_{345}} \, \langle6|k_3|4]\, (\epsilon_{61},k_3)  +
{s_{34}+s_{345}\over s_{34}-s_{345}} \, \langle6|k_5|4]\,
(\epsilon_{61},k_5) \right] \right]   \nonumber \\
& + & {[1\,5]\over [3\,5] }\, \left[
{\langle6\,3\rangle\,[3\,4]\over 2}\, \left(
(\lambda_5\tilde\lambda_3,\epsilon_{61}) -
 {(\lambda_5\tilde\lambda_3,k_4)\,(\epsilon_{61},k_5)
\over  s_{45}} \right) \right.  \nonumber \\
& + & \left.  {1\over9}\, s_{45}\,
(\tilde\epsilon_{45},\epsilon_{61}) - \tilde I_3^{2m(5)}(
\lambda_5\tilde\lambda_3,v_4) - \tilde I_3^{2m(5)}(
\lambda_5\tilde\lambda_3,\epsilon_{61},\lambda_6\tilde\lambda_4)
\right]
 \nonumber \\
& - & \langle2|k_3|4]\, \left[ {[1\,4]\over [5\,4]}\,
(I_3^{3m}(\lambda_6\tilde\lambda_2, \epsilon_{61}) - \tilde
I_3^{2m(4)}(\lambda_6\tilde\lambda_2, \epsilon_{61}) ) \right.
\nonumber \\
& + & \left. {[1\,5]\over [4\,5]}\,
I_4^{2mh(4)}(\lambda_5\tilde\lambda_4, \lambda_6\tilde\lambda_2,
\epsilon_{61})  -
I_4^{2me(2)}(\lambda_6\tilde\lambda_2,\lambda_5\tilde\lambda_1,
\epsilon_{61}) \right]  ,
\end{eqnarray}
where
\begin{eqnarray}
v_4 & = & (\lambda_6\tilde\lambda_4,k_3)\,\epsilon_{61} -
{1\over2}\, (\epsilon_6, \epsilon_1)\, \lambda_6\tilde\lambda_4 ,
\\ \tilde\epsilon_{45} & = & \epsilon_{45}|_{\epsilon_4 \to
 \lambda_6\tilde\lambda_4} =   \nonumber \\
& = & {\langle6\,5\rangle\over\langle4\,5\rangle}\,
\lambda_5\tilde\lambda_3 - {[3\,4]\over [5\,4]}\,
\lambda_6\tilde\lambda_4 + {\langle6\,5\rangle \, [3\,4] \over 2
\,  \langle 4\,5\rangle\, [5\,4]} \, (k_4-k_5) .
\end{eqnarray}

There are three different kinds of box diagrams shown in
Fig.~\ref{FeynmanSixPoint}. First, the rational part of the 2-mass-hard
box with two massless external lines $k_{2,3}$ is:
\begin{eqnarray}
R_3(1) & = &
I_4^{2mh(2)}(\epsilon_2,\epsilon_3,\epsilon_{45},\epsilon_{61};
-(\epsilon_4,\epsilon_5)/2, -(\epsilon_6,\epsilon_1)/2) .
\end{eqnarray}
The explicit expression in terms of the rational parts of triangle
integrals and other more primitive quantities is given in the
appendix.

The rational part of the two-mass-easy box with massless external
lines $k_{2,5}$ is decomposed into two parts $R_4(1) + R_4(4)$ and
we have:
\begin{eqnarray}
R_4(1)  & = & { \langle 6\,2\rangle [3\,5] \over \langle
5\,2\rangle [2\,5] }\, \left[ {
\langle5|k_{34}|2]^2(\epsilon_{34},k_2)(\epsilon_{61},k_2)\over 3
\,  (s_{34}-s_{234})\, (s_{61}-s_{345})} + {5\over18}\,
\langle5|\epsilon_{34}|2] \, \langle5|\epsilon_{61}|2]
\right] \nonumber \\
& + & {[3\,2] \over [5\,2] }\left[ \,
I_4^{2me(2)}(\epsilon_2,\epsilon_{34},k_5,\epsilon_{61})
\right. \nonumber \\
& + & {1\over4}\, \left.\left[ {(\epsilon_2,k_{34})\over
s_{34}-s_{234}} - {(\epsilon_2,k_{61})\over s_{61}-s_{345}}
\right] \, ((\epsilon_3,\epsilon_4)\epsilon_{61} +
(\epsilon_6,\epsilon_1)\epsilon_{34},k_2)
\right] \nonumber \\
& - & { \langle 6\,5\rangle [3\,2] \over  36 \, \langle
2\,5\rangle [5\,2] }\, ( 2 (k_2,k_5)\,
(\epsilon_{34},\epsilon_{61})  \nonumber \\
& &  - ( (\epsilon_{34},k_2)(\epsilon_{61},k_5)+
(\epsilon_{34},k_5)(\epsilon_{61},k_2)) ) .
\end{eqnarray}
$ R_4(4)$ is obtained from $ R_4(1)$ by the symmetry operation
$\tau^3$.  The rational part from one of the one-mass box diagrams,
i.e. diagram (e) with $i=2$ in Fig.~\ref{FeynmanSixPoint}, is:
\begin{eqnarray}
R_5(1) & = &  {\langle6\,2\rangle\over \langle4\,2\rangle }\, \left[
{4\over9}\, ( (\epsilon_3,k_2) (\epsilon_{561},k_4) +
(\epsilon_3,k_4) (\epsilon_{561},k_2) ) \right. \nonumber \\
& & - {5\over9}\, (k_2,k_4)\,
(\epsilon_3,\epsilon_{561})  + {1\over4}\,( (\epsilon_3,k_2) (\epsilon_{561},k_2)
+ (\epsilon_3,k_4) (\epsilon_{561},k_4) )  \nonumber \\
& + & \left.  {1\over4}\left[ {s_{34}+s_{234}\over
s_{34}-s_{234}}\, (\epsilon_3,k_2)\, (\epsilon_{561},k_2) +
{s_{23}+s_{234}\over s_{23}-s_{234}}\, (\epsilon_3,k_4)\,
(\epsilon_{561},k_4) \right] \right] \nonumber \\
& + & {\langle6\,4\rangle\over \langle2\,4\rangle }\, \left[  {1\over2}\,
((\epsilon_5,\epsilon_{61})+ (\epsilon_{56}, \epsilon_1)) \,
\tilde I_3^{2m(4)}(\epsilon_4,\epsilon_3) \right. \nonumber \\
& & \hskip 1cm \left.  + {1\over9}\, s_{34}\, (\epsilon_{34},\epsilon_{561}) - \tilde
I_3^{2m(4)}(\epsilon_4, \epsilon_{561},\epsilon_3) \right] .
\end{eqnarray}

The rational part for the three-mass triangle with external
momenta $(k_{23}, k_{45},k_{61})$, i.e. diagram (g) in
Fig.~\ref{FeynmanSixPoint}, is:
\begin{eqnarray}
R_0 & = &  I_3^{3m}(\epsilon_{23},\epsilon_{45},\epsilon_{61})
- {1\over 2}\, (\epsilon_2,\epsilon_3)\, I_3^{3m}(\epsilon_{45},\epsilon_{61})
\nonumber\\
& -&  {1\over 2}\, (\epsilon_4,\epsilon_5)\, I_3^{3m}(\epsilon_{61},\epsilon_{23})
- {1\over 2}\, (\epsilon_6,\epsilon_1)\, I_3^{3m}(\epsilon_{23},\epsilon_{45}) .
\end{eqnarray}

The rational part from the 2 two-mass triangle diagrams, i.e.
diagrams (h) and (i) with $i=2$ in Fig.~\ref{FeynmanSixPoint}, is:
\begin{eqnarray}
R_6(1) & = & I_3^{2m(2)}(\epsilon_2,\epsilon_{34}, \epsilon_{561}) -
{1\over 2}\,
I_3^{2m(2)}(\epsilon_2,v_6) \nonumber \\
& + & \tilde I_3^{2m(2)}(\epsilon_2,\epsilon_{345}, \epsilon_{61})
- {1\over 2}\,
\tilde I_3^{2m(2)}(\epsilon_2,\tilde v_6) \nonumber \\
v_6 & = & (\epsilon_3,\epsilon_4)\epsilon_{561} +
((\epsilon_5,\epsilon_{61}) + (\epsilon_{56},
\epsilon_1)) \, \epsilon_{34} , \\
\tilde v_6 & = &  ((\epsilon_3,\epsilon_{45}) +
(\epsilon_{34},\epsilon_5)) \, \epsilon_{61} +
(\epsilon_6,\epsilon_{1}) \, \epsilon_{345} .
\end{eqnarray}

All the one-mass triangle diagrams combined with the corresponding
bubble diagrams give a vanishing contribution to the rational part,
as all two neighboring gluons have opposite helicities for this
helicity configuration.  The remaining 3 bubble diagrams give the
following contribution to the rational part:
\begin{equation}
R_8 = -{1\over9} \, ( s_{123}\,(\epsilon_{123},\epsilon_{456}) +
s_{234}\,(\epsilon_{234},\epsilon_{561}) +
s_{345}\,(\epsilon_{345},\epsilon_{612}) ) .
\end{equation}

Adding all the above results with appropriate symmetry operations,
the final result for the rational part is:
\begin{equation}
R = F\,\left[ R_0 + ( {R_0}|_{i \to i+1})^*+ R_1 + R_8 +
\sum_{i=2}^6\sum_{j=1}^6 R_i(j)\right],
\end{equation}
where $R_i(j) = \tau^{j-1}( R_i(1) )$ and
\begin{equation}
F =  { 1\over \langle6\,2\rangle\,\langle2\,4\rangle\,\langle4\,6\rangle\,
[1\,5] \, [3\,1] \,  [5\,3] } ,
\end{equation}
is an overall factor.

\section{Factorization and the NMHV amplitudes}

The results obtained in previous 2 sections for the rational parts
of the 2 NMHV amplitudes are quite complicated analytic
expressions. To present the results in a readable form we have
used 3 levels of definitions. The first level consists of all the scalar products
(including all kinematic variables) in terms of spinor products.
The second level of definition consists of the composite polarization vectors
$\epsilon_{i(i+1)\cdots}$. The last level consists of the rational parts of
box and triangle integrals in terms of some elementary functions.

If we make a naive expansion of all the level 2 and 3 definitions,
the resulting analytic expression is actually quite large.
Although our results were derived rigorously from first principle,
it would be best if we can make some independent checks for these
results, especially to detect some human errors or typos. A
trivial but quite useful check is that the rational part should
have the correct spinor weight \cite{DixonReview}.

One stringent check is factorization. The factorization in
one-loop gauge theory was discussed in detail by Bern and Chalmers
in \cite{BernChalmers}. The 2-particle factorization properties
were discussed in \cite{BCDK,BDKY}. As two momenta become
collinear the one-loop amplitudes behave as \cite{BCDK,BDDK}
\begin{eqnarray}
A_n^{\rm {1-loop}} & \longrightarrow &  \sum_{\lambda=\pm}\left(
{\rm Split}_{-\lambda}^{\rm tree}(a^{\lambda_a}, b^{\lambda_b})\,
A_{n-1}^{\rm 1-loop}(\cdots (a+b)^{\lambda}\cdots) \right.
\nonumber \\
& & +\left . {\rm Split}_{-\lambda}^{\rm 1-loop}(a^{\lambda_a},
b^{\lambda_b})\, A_{n-1}^{\rm tree}(\cdots (a+b)^{\lambda}\cdots)
\right) ,
\end{eqnarray}
where $k_a\to z\, K$ and $k_b \to (1-z) \, K$, with $K=k_{a}+k_b$,
$K^2 = s_{ab} \to 0$. A similar relation also holds for
multi-particle factorization. The tree and 1-loop splitting
amplitudes in massless QCD have been given in
\cite{BDDK,BernChalmers,BDKM}. As we have a complete knowledge of
the 4- and 5-gluon amplitudes at tree and one-loop level, it is
straightforward to derive a set of relations which must be
satisfied by the 6-gluon amplitude. These relations provide an
extremely stringent check.

Berger, Bern, Dixon, Forde and Kosower have done this check by
using their factorization check program and found that the results
given in the previous two sections in fact pass the check for all
channels. The Mathematica codes used by them were input by us by
using the formulas given in this paper. The codes are available up
to requests and/or they will be stored somewhere for free use.

\section{Conclusion}

In this paper, by using the formalism developed in a previous
paper \cite{xyzi}, we computed explicitly the rational parts
 of the 6-gluon one-loop amplitudes for all the possible helicity
 configurations. In particular we present most of the intermediate
results for the four helicity configurations which have no available
 published results in the literature. For
two of the MHV helicity configurations our results agree with the all multiplicity
results \cite{BDKE} which are
obtained by using the bootstrap recursive approach of Bern,
Dixon and Kosower \cite{BDKA,BDKB}.  For other two non-MHV  helicity configurations our results
passed the factorization tests in all channels. By combining our results with the
previously computed
cut-constructible parts \cite{BoFengSix}, we have all the ingredients for a complete
expression of the 6-gloun one-loop  QCD
 amplitude. This paves the way for assembling new NLO helicity amplitudes in
numerical codes for cross-section calculations.

The method presented in this paper can be applied straightforwardly to compute all
the rational
parts of the 6-parton one-loop QCD amplitudes. Apart from the rational parts of the
3-mass
tensor box integrals, our method can also be applied to compute the rational parts
for $e^+e^-\to 5$-partons.
As we demonstrated in this paper all the intermediate steps of our computations are
carried out by hand.
Only for the input of composite polarization vectors and for combining all the
results together (by using
various permutation symmetry operations) we  used computer symbolic system
Mathematica.
There should be no much difficulty to automate our method (to compute the rational
parts).
In this way we may push the limit to 8-gluon or higher.

\section*{Appendix: Formulas for the rational parts of some Feynman
integrals}

For quick reference we list here the explicit formulas for the
rational parts of some Feynman integrals. The derivation can be
found in \cite{xyzi}.

Firstly, for the bubble integral we have:
\begin{eqnarray}
I_2(\epsilon_1,\epsilon_2) & = & \int { {\rm d}^D p \over i \pi^{D/2} } \, {
(\epsilon_1,p)(\epsilon_2,p) \over p^2 ( p+K)^2} \nonumber \\
& = & {1\over 18}\, ( (\epsilon_1,K)\, (\epsilon_2,K) -  2 \,
K^2\, (\epsilon_1,\epsilon_2) ) .
\end{eqnarray}
where $K$ is the sum of momenta on one side of the bubble diagram.

For three-mass triangle
integral with external momenta $\{K_1,K_2,K_3\}$ and arbitrary
polarization vectors $\epsilon_{1,2,3}$  we have
\begin{eqnarray}
I_3(\epsilon_1, \epsilon_2, \epsilon_3 ) & \equiv &  \int { {\rm
d}^D p \over i \pi^{D/2} } \, {  (\epsilon_1 ,p)  \, (\epsilon_2 ,
 p-K_1)  \,
(\epsilon_3 ,p+K_3)  \over p^2 (p-K_1)^2 (p+K_3)^2 }  , \nonumber
\\
& = & -
  F_0(s_1, s_2, s_3) ( (\epsilon_1, K_1) \, (\epsilon_2,
K_1) \, (\epsilon_3, K_2)   \nonumber \\
& &   + (\epsilon_1, K_3) \, (\epsilon_2, K_2)\, (\epsilon_3, K_2)
+(\epsilon_1, K_3) \, (\epsilon_2, K_1)
\, (\epsilon_3, K_3) \nonumber \\
& &    +(\epsilon_1, K_3) \, (\epsilon_2, K_1)\,(\epsilon_3, K_2
)-(\epsilon_1, K_1)\, (\epsilon_2, K_2)\, (
\epsilon_3, K_3)  ) \nonumber \\
& - &   \sum_{i=1}^3 (\epsilon_1, K_i)\, (\epsilon_2, K_i)\,
(\epsilon_3, K_ i)  \,  F_i(s_1, s_2, s_3)
\nonumber \\
& - &  {1\over 2 \, \Delta} \Big( (s_1 - s_2 - s_3) \,
(\epsilon_1, K_1) \, (\epsilon_2, K_1) \, (\epsilon_3, K_3)
\nonumber \\
& & + \, (s_2 - s_3 - s_1) \, (\epsilon_1, K_1) \, (\epsilon_2,
K_2) \,  (\epsilon_3, K_2)
\nonumber \\
& & + \, (s_3 - s_1 - s_2) \, (\epsilon_1, K_3) \, (\epsilon_2,
K_2) \, (\epsilon_3, K_3)  \Big) \nonumber \\
& + &  {7 \over 36} \, \Big( (\epsilon_1, \epsilon_2)\,
(\epsilon_3, K_3 - K_2) +
 (\epsilon_2, \epsilon_3)\, (\epsilon_1, K_1 - K_3) \nonumber \\ & & +
 (\epsilon_3, \epsilon_1)\, (\epsilon_2, K_2 - K_1)  \Big) \nonumber \\
 & + & {1\over 12\, \Delta} \, \Big( (\epsilon_1, \epsilon_2)\,
(\epsilon_3, K_3 - K_2)\, { s_1(s_2 + s_3 - s_1)} \nonumber \\ & &
+
 (\epsilon_2, \epsilon_3)\, (\epsilon_1, K_1 - K_3)\, { s_2(s_3 + s_1 - s_2) } \nonumber \\ & &
 +
 (\epsilon_3, \epsilon_1)\, (\epsilon_2, K_2 - K_1)\, { s_3(s_1 + s_2 -
s_3)} \Big)
 \nonumber \\
& + & {1\over 12\, \Delta} \, \Big( (\epsilon_1, \epsilon_2)\,
(\epsilon_3, K_1) \, {(s_3-s_2) (s_2+s_3 - s_1) } \nonumber \\ & &
+
 (\epsilon_2, \epsilon_3)\, (\epsilon_1, K_2)\, {(s_1-s_3) (s_3+s_1 - s_2)}\nonumber \\ & & +
 (\epsilon_3, \epsilon_1)\, (\epsilon_2, K_3)\, {(s_2-s_1) (s_1+s_2 - s_3)} \Big)
 ,
\end{eqnarray}
where
\begin{eqnarray}
F_0(s_1,s_2,s_3) & = & {10 \, s_1 s_2 s_3 \over 3 \Delta^2 } + {
(s_1 + s_2  + s_3) \over 6 \Delta}, \\
F_1(s_1,s_2,s_3) & = & {5 \, (s_1 + s_2 - s_3)\, s_2 s_3  \over 3
\Delta^2 } + {(s_1 - s_3) \over 3 \Delta}, \\
F_2(s_1,s_2,s_3) & = & {5 \, (s_2 + s_3 - s_1)\, s_3 s_1  \over 3
\Delta^2 } + {(s_2 - s_1) \over 3 \Delta}, \\
F_3(s_1,s_2,s_3) & = & {5 \, (s_3 + s_1 - s_2)\, s_1 s_2  \over 3
\Delta^2 } + {(s_3 - s_2) \over 3 \Delta} , \\
\Delta & = & s_1^2 + s_2^2 + s_3^2 - 2 \,( s_1\,s_2 + s_2\, s_3+
s_3\,s_1),
\end{eqnarray}
and $s_i = K_i^2$.

For degree 2 polynomial we have:
\begin{eqnarray}
I_3 (\epsilon_i,\epsilon_j ) & \equiv &  \int { {\rm d}^D p \over
i \pi^{D/2} } \, {   (\epsilon_i,p) \,  (\epsilon_j,p) \over p^2
(p-K_1)^2 (p+K_3)^2 } \nonumber \\
& = & - {1\over 2\, \Delta} \, \Big( s_1 \, ((\epsilon_i , K_2) \,
(\epsilon_j , K_3)
+ (\epsilon_i , K_3) \, (\epsilon_j , K_2)) \nonumber \\
&   + &   s_2 \, ((\epsilon_i , K_3) \, (\epsilon_j , K_1)+
(\epsilon_i , K_1) \, (\epsilon_j , K_3)) \nonumber \\
&   + & s_3 \, ((\epsilon_i , K_1) \, (\epsilon_j , K_2) +
(\epsilon_i , K_2) \, (\epsilon_j , K_1))  \Big)  + {1\over 2} \,
(\epsilon_i,\epsilon_j)  .
\end{eqnarray}

For 2-mass triangle integrals the above formulas simplify greatly.
They are:
\begin{eqnarray}
I_3 (\epsilon_1,\epsilon_2) & \equiv &   \int {{\rm d}^D p \over i
\pi^{D/2}} \,
  { (\epsilon_1, p) \, (\epsilon_2 , p) \,  \over p^2(p-k_1)^2
  (p+K_3)^2}    \nonumber \\
 & = &      {1\over 2}\, (\epsilon_1, \epsilon_2) +
   {(K^2_2 + K^2_3)\over 2 (K^2_2-K^2_3)^2 }
  \, (\epsilon_1, k_1) \, (\epsilon_2, k_1)  \nonumber \\
& + & {( (\epsilon_1, K_2)\,(\epsilon_2,k_1)  -(\epsilon_1, k_1)\,
(\epsilon_2, K_3) ) \over 2(K^2_2-K^2_3)}
 , \label{twomassone} \\
I_3 (\epsilon_1,\epsilon_2) & =  &
     {1\over2}\, (\epsilon_1, \epsilon_2) +  { (\epsilon_1, K_2) \,
(\epsilon_2, k_1)  \over 2(K_2^2-K^2_3)}  , \qquad (\epsilon_1,
k_1) = 0, \label{twomasstwo}
\end{eqnarray}
and
\begin{eqnarray}
I_3 (\epsilon_i) & \equiv &  \int {{\rm d}^D p \over i \pi^{D/2}}
\,\,
  { (\epsilon_1, p)  \, (\epsilon_2 ,  p-k_1) \, (\epsilon_3, p) \,
 \over p^2\, (p-k_1)^2 \,   (p+K_3)^2}
  \nonumber \\
  &   =  &   {1\over 36} \Big( (\epsilon_2,  4\,K_2 -7\, k_1)\,(\epsilon_1,
\epsilon_3) -(2 \leftrightarrow 3)  + 4 (\epsilon_1, K_2)\,
(\epsilon_2, \epsilon_3) \Big)  \nonumber \\
& - & {(K^2_2 + K^2_3)  \over 6\,(K^2_2-K^2_3)^2 } \, (\epsilon_1,
K_2) \,   (\epsilon_2, k_1) \, (\epsilon_3, k_1)
\nonumber \\
 & - & {(\epsilon_1, K_2)\, ((\epsilon_2, k_1)\,
(\epsilon_3, K_3) - (\epsilon_2 , K_2)\, (\epsilon_3, k_1))\over
6\, (K^2_2-K^2_3)}
\nonumber \\
& - & { (K^2_2 + K^2_3) \over 12\, (K^2_2-K^2_3) } \, (
(\epsilon_1, \epsilon_2)\, (\epsilon_3, k_1) + (\epsilon_1 ,
\epsilon_3)\, (\epsilon_2, k_1) )  , \label{twomassthree}
\end{eqnarray}
where $\epsilon_1$ satisfies the physical condition
$(\epsilon_1,k_1)$ = 0 and $\epsilon_{2,3}$ are arbitrary
4-dimensional polarization vectors.

For degree 3 two-mass-easy box integrals we have:
\begin{eqnarray}
I_4^{2me} (\lambda_3\tilde\lambda_1,\epsilon_2, \epsilon_3)  =
 {\langle 3 |K_2|1 ]   \over 2} \, \left[ { (\epsilon_2 , k_3) \,
(\epsilon_3 , k_3) \over (K_2^2 - t) (K_4^2 - s) } - (k_3
\rightarrow k_1, s \leftrightarrow t) \right]  .
\end{eqnarray}
If two of the polarization vectors satisfy the physical condition, i.e. $(\epsilon_1,k_1)=0$
and $(\epsilon_3,k_3)=0$, we have
\begin{eqnarray}
I_4^{2me} (\epsilon_1,\epsilon_2, \epsilon_3) & = &
 - {(\epsilon_1,k_3)\,(\epsilon_3,k_1)\over 2 \, (k_1,k_3) } \,\left[
 {(\epsilon_2,k_3)\over K_2^2 - t } + {(\epsilon_2,k_1)\over K_4^2 -t} \right]
\nonumber \\
&   & \hskip -1cm - { (\epsilon_1,K_2)
\,(\epsilon_2,k_1)\,(\epsilon_3,k_1) \over 2 \,  (K_2^2 -
s)\,(K_4^2 - t)} - { (\epsilon_1,k_3)
\,(\epsilon_2,k_3)\,(\epsilon_3,K_4) \over 2 \,  (K_2^2 -
t)\,(K_4^2 - s)} .
\end{eqnarray}
The formulas for the degree 3 two-mass-hard box integrals are:
\begin{eqnarray}
I_4^{2mh}(\lambda_1\tilde\lambda_2, \epsilon_2,\epsilon_3) & = & {
\langle 1 | K_3 | 2 ]
\over 4 \, \delta } \,  I_4(\epsilon_2,\epsilon_3) , \\
I_4^{2mh}(\lambda_2\tilde\lambda_1, \epsilon_2,\epsilon_3) & = & {
\langle 2 | K_3 | 1 ] \over 4 \, \delta } \,
I_4(\epsilon_2,\epsilon_3) ,
\end{eqnarray}
where
\begin{eqnarray}
I_4(\epsilon_2,\epsilon_3)  & = & (\epsilon_2, k_1)(\epsilon_3,
K_4) + (\epsilon_2, K_4)(\epsilon_3, k_1) +
 (\epsilon_2, k_2) (\epsilon_3, K_3) \nonumber \\
 & + &  (\epsilon_2, K_3)(\epsilon_3, k_2)
 - { 1\over   \Delta} \left[  2 \, ( K_3^2 K_4^2 - t^2  + \delta )
(\epsilon_2, k_{12}) (\epsilon_3 , k_{12})  \right. \nonumber \\
& + &  (K_3^2 + K_4^2 - s - 2\, t)   \, ( (K_3^2 - K_4^2 + s)
(\epsilon_2 , K_4) (\epsilon_3 , K_4) \nonumber \\
& & \left.  + (K_4^2 - K_3^2 + s) \,
(\epsilon_2 , K_3) (\epsilon_3 , K_3) ) \right] \nonumber \\
& + & {K_4^2 + t \over K_4^2 - t} \, { (\epsilon_2 , k_1)
(\epsilon_3 , k_1)  } + {K_3^2 + t \over K_3^2 - t} \, {
(\epsilon_2 , k_2) (\epsilon_3 , k_2)  } ,
\end{eqnarray}
and
\begin{eqnarray}
 \delta & = & K_3^2 \, K_4^2 - (K_3^2 + K_4^2 )\, t + (s + t)\, t, \\
\Delta  & = & \Delta(k_{12}^2,K_3^2,K_4^2),  \\
\Delta(s_1,s_2,s_3) & = & s_1^2 + s_2^2 + s_3^2 - 2(s_1s_2+ s_2s_3
+ s_3s_1),
\end{eqnarray}
are functions of the external momentum invariants.

In order to give the formulas for the rational parts of the degree
4 polynomials, we define:
\begin{equation}
I_4 (\epsilon_1,\epsilon_2, \epsilon_3, \epsilon_4)   \equiv
 \int { {\rm d}^D p \over i \pi^{D/2}}
\, { (\epsilon_1, p) \, (\epsilon_2 ,  p-K_1) \, (\epsilon_3 ,  p
- K_{12})\, (\epsilon_4, p+K_4)  \over p^2(p-K_1)^2 (p-K_{12})^2
(p+K_4)^2}  ,
\end{equation}
where $K_{12} = K_1 + K_2$. For two-mass-easy cases
($K_{i,3}=k_{1,3}$ are massless external lines), we have
\begin{eqnarray}
 I_4(\lambda_3\tilde\lambda_1, \epsilon_2,\lambda_1\tilde\lambda_3,
 \epsilon_4)  & = &  -  {1\over 4} \left( {K_2^2 + s \over K_2^2 - s } +
 {K_4^2 + t \over K_4^2 - t }\right) \,(\epsilon_2,k_1)(\epsilon_4,k_1)
 \nonumber \\
&   & \hskip -1.5cm
 - {1\over 4} \left( {K_2^2 + t \over K_2^2 - t
} +  {K_4^2 + s \over K_4^2 - s }\right)
\,(\epsilon_2,k_3)(\epsilon_4,k_3)
-  {5\over 9} \, (k_1,k_3)( \epsilon_2,  \epsilon_4) \nonumber \\
& +  &  {4\over 9}\, \Big( (\epsilon_2,k_1)(\epsilon_4,k_3) +
 (\epsilon_2,k_3)(\epsilon_4,k_1) \Big)   , \\
I_4(\lambda_1\tilde\lambda_3, \epsilon_2,\lambda_1\tilde\lambda_3,
 \epsilon_4)  & = &  {5\over 9}\, \langle 1|\epsilon_2|3] \, \langle1|\epsilon_4|3]
 \nonumber \\
 &  & \hskip -2cm + {\langle1|K_2|3]^2 \over 3}\, \left[
 {(\epsilon_2,k_1)\,(\epsilon_4,k_1) \over (K_2^2 - s) \, (K_4^2 - t)} +
 {(\epsilon_2,k_3)\,(\epsilon_4,k_3) \over (K_2^2 - t) \, (K_4^2 - s)} \right]   .
\end{eqnarray}
Other cases can be either obtained by conjugation or relabelling
$k_{1,3}$.

For 2-mass-hard box cases ($K_{1,2} =k_{1,2}$ are two massless external
lines),  we define:
\begin{eqnarray}
I_4^{2mh}(\epsilon_1,\epsilon_2,\epsilon_3,\epsilon_4;c_3,c_4)  & &   \nonumber \\
&   & \hskip -4cm  \equiv  I_4[ (\epsilon_1,p)
(\epsilon_2,p-k_1)((\epsilon_3, p+K_4)+c_3)((\epsilon_4, p+K_4)+c_4)] \nonumber \\
&    & \hskip -4cm  = \int{ {\rm d}^Dp \over i \pi^{D/2}} \, {
(\epsilon_1,p) (\epsilon_2,p-k_1)((\epsilon_3,
p+K_4)+c_3)((\epsilon_4, p+K_4)+c_4) \over p^2\, (p-k_1)^2\,
(p-k_{12})^2\, (p+K_4)^2 } , \nonumber \\
\end{eqnarray}
and
\begin{eqnarray}
& & \hskip -1.2cm
I_4^{2mh}(\lambda_1\tilde\eta_1,\eta_2\tilde\lambda_2,
\epsilon_3,\epsilon_4;c_3,c_4)
  \nonumber \\
& = & {1 \over \langle  2| K_3 |1]} \Big[ - \frac{1}{6} \,
\langle\eta_2| k_2K_3k_1|\tilde\eta_1\rangle \,
(\epsilon_3,\epsilon_4) - t\,\langle\eta_2\,2\rangle \,
[\tilde\eta_1\,1] \,
I_4^{2mh}(\lambda_1\tilde\lambda_2,\epsilon_3,\epsilon_4)
\nonumber \\
& + &  t \, \langle\eta_2|k_2|\tilde\eta_1 ] \,
I_3^{3m}(\epsilon_3,\epsilon_4) + \langle\eta_2\,2\rangle \,
[\tilde\eta_1\,1] \left( {1\over 2}
(\langle1|\epsilon_3|2] \, c_4 + \langle1|\epsilon_4|2] \, c_3) \right. \nonumber \\
& & \left. + {1\over 18} (\langle1|\epsilon_3|2] \, \epsilon_4 +
\langle1|\epsilon_4|2] \,
\epsilon_3, 7\,k_1+2\,k_2 + 9\,K_4)\right) \nonumber \\
& + & {1\over18}\, \langle  \eta_2\,2\rangle \,[\tilde\eta_1\,2]
\, ( (\epsilon_3,k_{12})\,(\epsilon_4,k_{12}) - 2 \, s_{12} \,
(\epsilon_3,\epsilon_4) ) \nonumber \\
  & + & \frac{1}{18}\, \langle\eta_2|(k_2+K_3)|\tilde\eta_1] (
(\epsilon_3, k_2+K_3)(\epsilon_4,
 k_2+ K_3) - 2 \, t\, (\epsilon_3, \epsilon_4) ) \nonumber \\
& - & \frac{1}{18}\, \langle\eta_2|K_3|\tilde\eta_1] \, (
(\epsilon_3, K_3)(\epsilon_4, K_3) - 2
K_3^2(\epsilon_3,\epsilon_4)) \nonumber \\
 & +  & \langle
2|K_3|\tilde\eta_1\rangle ( I_3^{2m}(\eta_2\tilde\lambda_2,
\epsilon_3,\epsilon_4) \nonumber \\
& + & I_3^{2m} (\eta_2\tilde\lambda_2,
(c_3-(\epsilon_3,K_3))\epsilon_4 + (c_4
+ (\epsilon_4,K_4+k_1)) \epsilon_3) ) \nonumber \\
& + &  I_3^{3m}(v, \epsilon_3,\epsilon_4) +  I_3^{3m}(v,
(c_3-(\epsilon_3,K_3))\epsilon_4 + c_4\epsilon_3 ), \\
 & + &  \langle\eta_2|K_4|1]\, ( \tilde I_3^{2m}(\lambda_1\tilde\eta_1,\epsilon_3,
\epsilon_4)\nonumber \\
& + &\tilde
I_3^{2m}(\lambda_1\tilde\eta_1,(c_3-(\epsilon_3,k_2+K_3))\epsilon_4+
(c_4+(\epsilon_4,K_4))\epsilon_3 ) \Big],
\end{eqnarray}
where
\begin{equation}
v =  \langle\eta_2|K_3|1] \lambda_1\tilde\eta_1 +
\langle\eta_2|K_3|2] \lambda_2\tilde\eta_1
-(k_2,K_3)\eta_2\tilde\eta_1 .
\end{equation}
The opposite helicity case is obtained by conjugation.

\section*{Acknowledgments}

We would like to thank Carola F. Berger, Zvi Bern, Lance J. Dixon,
Darren Forde and David A. Kosower for sending us their results
prior to publication \cite{BDKE}, and for discussions, reading the
paper, comments and assistance in comparing our six-gluon results
with theirs. We also thank them for checking the NMHV results by
using their factorization check program.  CJZ would like to thank
J. -P. Ma for constant encouragements, helpful discussions and
careful reading of the paper. His (financial) support (to buy a
computer which was still in use today) actually goes back to the
much earlier difficult times when I did not have enough grants from
other sources. What is more important is that there are no strings
attached to his support and it is up to the last author to explore
what he wants to. CJZ would also like to thank R. Iengo for
encouragements and his interests in this work, helpful discussions
and comments; to Z.~Chang, B.~Feng, E.~Gava, H.~-Y.~Guo,
K.~S.~Narain, K.~Wu, Y.~-S.~Wu, Z.~Xu and Z.~-X.~Zhang for
discussions and comments; to Prof.~X.~-Q.~Li and the hospitality
at Nankai University where we can have good food; and finally to
Prof.~S.~Randjbar-Daemi and the hospitality at Abdus Salam
International Center for Theoretical Physics, Trieste, Italy. This
work is supported in part by funds from the National Natural
Science Foundation of China with grant number 10475104 and
10525522.

\end{document}